\documentclass[12pt]{article}
\usepackage{epsfig,amssymb,amsmath,psfrag,subfigure,rotate,color,wasysym}

\allowdisplaybreaks
\newcommand{\insertfig}[2]{\includegraphics[width=#1cm]{#2}}

\DeclareMathOperator*{\SumInt}{%
\mathchoice%
  {\ooalign{$\displaystyle\sum$\cr\hidewidth$\displaystyle\int$\hidewidth\cr}}
  {\ooalign{\raisebox{.14\height}{\scalebox{.7}{$\textstyle\sum$}}\cr\hidewidth$\textstyle\int$\hidewidth\cr}}
  {\ooalign{\raisebox{.2\height}{\scalebox{.6}{$\scriptstyle\sum$}}\cr$\scriptstyle\int$\cr}}
  {\ooalign{\raisebox{.2\height}{\scalebox{.6}{$\scriptstyle\sum$}}\cr$\scriptstyle\int$\cr}}
}

\def\XXint#1#2#3{{\setbox0=\hbox{$#1{#2#3}{\int}$ }
\vcenter{\hbox{$#2#3$ }}\kern-.6\wd0}}

\def \be  {\begin{equation}}
\def \ee  {\end{equation}}
\def \ba  {\begin{eqnarray}}
\def \ea  {\end{eqnarray}}
\def \baa {\begin{eqnarray*}}
\def \eaa {\end{eqnarray*}}
\def \lab #1 {\label{#1}}

\newcommand\re[1]{(\ref{#1})}
\def\d{\hbox{{d}\kern-.20em\hbox{l}}}

\def \matrix #1 {\left(\begin{array}{cc} #1 \end{array}\right)}

\def \res{\mathop{\rm res}\nolimits}

\newcommand \vev [1] {\langle{#1}\rangle}

\newcommand{\bit}[1]{\mbox{\boldmath$#1$}}

\newcommand{\ft}[2]{{\textstyle\frac{#1}{#2}}}
\numberwithin{equation}{section}
\begin{document}

\begin{titlepage}

\thispagestyle{empty}

\vspace*{3cm}

\centerline{\large \bf Resummed tree heptagon}
\vspace*{1cm}

\centerline{\sc A.V.~Belitsky}

\vspace{10mm}

\centerline{\it Department of Physics, Arizona State University}
\centerline{\it Tempe, AZ 85287-1504, USA}

\vspace{2cm}

\centerline{\bf Abstract}

\vspace{5mm}

The form factor program for the regularized space-time S-matrix in planar maximally supersymmetric gauge theory, known as the pentagon operator product expansion, 
is formulated in terms of flux-tube excitations propagating on a dual two-dimensional world-sheet, whose dynamics is known exactly as a function of 't Hooft 
coupling. Both MHV and non-MHV amplitudes are described in a uniform, systematic fashion within this framework, with the difference between the two encoded in 
coupling-dependent helicity form factors expressed via Zhukowski variables. The nontrivial  SU(4) tensor structure of flux-tube transitions is coupling independent and 
is known for any number of charged excitations from solutions of a system of Watson and Mirror equations. This description allows one to resum the infinite 
series of form factors and recover the space-time S-matrix exactly in kinematical variables at a given order of perturbation series. Recently, this was done for the hexagon.
Presently, we successfully perform resummation for the seven-leg tree NMHV amplitude. To this end, we construct the flux-tube integrands of the fifteen independent Grassmann 
component of the heptagon with an infinite number of small fermion-antifermion pairs accounted for in NMHV two-channel conformal blocks.

\end{titlepage}

\setcounter{footnote} 0

\newpage

%\pagestyle{plain}
%\setcounter{page} 1

%{
%\footnotesize 
%\tableofcontents}

%\newpage

\section{Introduction}

The duality between the space-time scattering matrix and supersymmetric Wilson loop on a null polygonal contour 
\cite{Alday:2007hr,Drummond:2007cf,Brandhuber:2007yx,CaronHuot:2010ek,Mason:2010yk,Belitsky:2011zm} in planar $\mathcal{N}=4$ superYang-Mills
theory was instrumental in formulating of a non-perturbative framework for the former in terms of two-dimensional physics taking central stage in the latter. The
dynamics of excitations propagating on the corresponding background is exactly solvable and allows one to determine their dispersion relations and scattering 
matrices at any value of 't Hooft coupling. This current approach \cite{Basso:2013vsa} to scattering amplitudes emerged from the study of the near-collinear expansion 
of Wilson loop expectation value \cite{Alday:2010ku,Gaiotto:2011dt} when two adjacent links merging at a cusp tend to straighten up. Deviation from the straight line admits a s
ystematic expansion in a series of operator insertions into the gauge link. These operators create the aforementioned excitations of the flux-tube stretched between 
the Wilson loop contour. They propagate on the two-dimensional world-sheet and get absorbed via a mechanism analogous to their creation. At any order in the 
power of the deviation parameter there is a finite number of contributing particles, which however have to be summer over in order to get the exact representation 
of the Wilson loop and correspondingly space-time scattering amplitudes in generic kinematics.

The series representation of the $n$-gon superWilson loop \cite{Basso:2013vsa} 
\begin{align}
\label{MatrixElementSuperPentagon}
\mathbb{W}_n
=
\SumInt_{N,N', \dots, N''}
\vev{0 | \mathbb{P}_{n-4} | {\rm\bf p}_{N''} (\bit{u}'')}
\dots
\vev{ {\rm\bf p}_{N'} (\bit{u}') | \mathbb{P}_2 | {\rm\bf p}_N (\bit{u})}
\vev{ {\rm\bf p}_N (\bit{u}) | \mathbb{P}_1 | 0 }
\, ,
\end{align}
is given in terms of the creation/annihilation/transition form factors $\vev{ {\rm\bf p}_{N'} (\bit{u}') | \mathbb{P} | {\rm\bf p}_N (\bit{u})}$ of pentagon operators $\mathbb{P}$ 
between the states of the flux-tube with rapidities $\bit{u} = (u_1, \dots, u_N)$ and $\bit{u}' = (u'_1, \dots, u'_{N'})$, which are integrated over
\begin{align}
\int_N \to
\int \prod_{i=1}^N \frac{d u_i}{2 \pi} \mu_{{\rm p}_i} (u_i) {\rm e}^{- \tau E_{{\rm p}_i} (u_i) + i \sigma p_{{\rm p}_i} (u_i) + i h_{{\rm p}_i} \varphi}
\, ,
\end{align}
with measures $\mu_{{\rm p}_i}$ and propagation phases determined by their energy $E_{{\rm p}_i}$, momentum $p_{{\rm p}_i}$ and helicity $h_{{\rm p}_i} $ accompanied 
by the reciprocal variables of time $\tau$, space $\sigma$ and rotation angle $\varphi$. In the above formula, each $N$ of the intermediate state cumulatively denotes the 
number of particles, their helicities and SU(4) quantum numbers in the completeness condition. The form factor decompose
\begin{align}
\label{PentagonFF}
\vev{ {\rm\bf p}_{N'} (\bit{u}') | \mathbb{P} | {\rm\bf p}_N (\bit{u})}
=
\pi^{\bf\scriptscriptstyle R} (\bit{u}|\bit{u}') [ \Pi^{\bf\scriptscriptstyle R} ]_{N|N'} (\bit{u} | \bit{u}') P (\bit{u}|\bit{u}')
\, ,
\end{align}
as a product of the coupling-constant dependent dynamical $P (\bit{u}|\bit{u}')$ and helicity form factors $\pi^{\bf\scriptscriptstyle R} (\bit{u}|\bit{u}')$, with the latter depending 
on the SU(4) charge ${\rm\bf R}$ of the pentagon inducing a given transition, and the coupling-independent tensor $[ \Pi^{\bf\scriptscriptstyle R} ]_{N|N'} (\bit{u} | \bit{u}')$ 
carrying representation indices of the flux-tube excitations. While, the helicity form factors are simply given by the product of individual single-particle form factors,
which in turn are powers of the Zhukowski variables, the expression for the dynamical part $P (\bit{u}|\bit{u}')$, while factorizable, takes on a more complicated form.
It reads
\begin{align}
\label{FactorizedPentagon}
P (\bit{u} | \bit{u}')
=
\frac{\prod_{i = 1}^{N} \prod_{j=1}^{N'} P (u_i|u'_j)}{\prod_{i>j}^{N} P (u_i|u_j) \prod_{k<l}^{N'} P (u'_k|u'_l)}
\, ,
\end{align}
in terms of one-to-one particle transition form factors, where for brevity, we do not display flavors of the excitations involved. All of the ingredients on the right-hand side of
Eq.\ \re{PentagonFF} are known from a series of papers 
\cite{Basso:2013vsa,Basso:2013aha,Basso:2014koa,Belitsky:2014sla,Basso:2014nra,Belitsky:2014lta,Belitsky:2015efa,Basso:2014hfa,Basso:2015rta,Belitsky:2016vyq}.

The resummation of the entire form factor series is by no means obvious. Recently, it was successfully accomplished in Ref.\ \cite{Cordova:2016woh} for the simplest 
case of the hexagon\footnote{Previous attempts include weak coupling analyses of a double scaling limit relevant for multi-Regge regime \cite{Drummond:2015jea} 
and resummations at strong coupling \cite{Fioravanti:2015dma,Belitsky:2015qla,Bonini:2015lfr} which yield TBA expectations and beyond 
\cite{Basso:2014jfa,Belitsky:2015lzw,Bonini:2017gwt}.} making use of the available integral representation for the traced SU(4) tensor structures \cite{Basso:2015uxa}. 
This was based on the notion of an effective particle. The latter is built from an elementary scalar $\phi$, large (anti)fermion $(\bar\psi) \psi$, (anti)gluon $(\bar{g}) g$ 
excitations and bound states $(\bar{g}_a) g_a$ carrying the intrinsic quantum numbers and a cloud of small fermions and antifermions. In fact, this picture is alike 
the traditional constituent quark model, where the constituent excitation carries the quantum numbers of the current quark surrounded by an un-obscuring cloud of glue 
and quark-antiquark pairs, which merely renormalizes its mass. The observation that the number of effective excitations grown slowly with each order of perturbation 
theory, allows one to operate in terms of a very small number of constituent particles at lowest loop orders.

In this paper, we extend the program to reconstruction of the NMHV heptagon in full kinematics from the form factor expansion making extensive use of the SU(4) tensor
part following Ref.\ \cite{Belitsky:2016vyq}. This will be done to leading order in 't Hooft coupling, i.e., tree level. The subsequent presentation is organized as follows. In 
the next section, we will recall the parent excitations with various SU(4) quantum numbers that will contribute to the Grassmann components of the heptagon. Then, 
in Sect.~\ref{SectionIndepHeptagonComp}, after briefly reminding the structure of the NMHV heptagon, we provide results for its fifteen independent components in terms 
of the effective excitations. Next, we turn to the resummation of small fermion-antifermion pairs, which determines as a result the NMHV conformal blocks. Finally, we conclude. 
A few appendices contain calculational details on polygon kinematics and reconstruction of charged SU(4) tensors from singlet ones.

%%%%%%%%%%%%%%%%%%%%%%%%%%%%%%%%%%%%%%%%%%%%%%%%%%%%%%%%%%%%%%%%%%%%%
%            Figure
%%%%%%%%%%%%%%%%%%%%%%%%%%%%%%%%%%%%%%%%%%%%%%%%%%%%%%%%%%%%%%%%%%%%%
\begin{figure}[t]
\begin{center}
\mbox{
\begin{picture}(0,200)(60,0)
\put(0,-140){\insertfig{16}{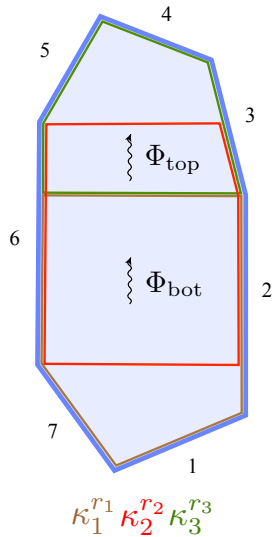}}
\end{picture}
}
\end{center}
\caption{ \label{Heptagon} Pentagon decomposition of the NMHV heptagon.
}
\end{figure}
%%%%%%%%%%%%%%%%%%%%%%%%%%%%%%%%%%%%%%%%%%%%%%%%%%%%%%%%%%%%%%%%%%%%%

\section{Parent excitations}
\label{SectionExcitations}

The fundamental flux-tube excitations consist of the SU(4) singlet gluon and antigluon, i.e., opposite helicity $\pm 1$ states, fermion and antifermion in the fundamental
representation of dimension ${\bf 4}$ and its conjugate ${\bf \bar{4}}$, respectively, and antisymmetric ${\bf 6}$ scalars. The (anti)gluons form bound states of
helicity $|h| >1$. This is not the only way to form a parent excitation which is the lowest weight for a tower of descendants to build an effective one. One can encode 
the same quantum numbers by forming strings in rapidities with (anti)fermions. This comes about from the following unique feature of the flux-tube (anti)fermion, its natural 
rapidity is the Zhukowski variable $x$ rather than the ``bare" rapidity $u = x + g^2/x$. As a consequence two copies of the $u$-plane are needed to cover the entire 
$x$-plane. The two complex planes are glued across the cut $[-2g, 2g]$. The fermions with $|x| > g$, live on the top sheet and have large rapidities $x \sim u$, while for 
$|x| < g$, they possess small rapidities $x \sim g^2/u$. They are called the large and small fermion, respectively. The latter can reach zero momentum (or which is equivalent, 
infinite rapidity) where it becomes a generator of the supersymmetric transformation. Thus, one can glue as many (anti)fermions to a given fundamental excitation without 
any cost in 't Hooft coupling as long as cumulative quantum numbers allow it.

Namely, for a given dimension-${\bf R}$ representation of SU(4), we decompose parent excitations into positive and negative helicities
\begin{align}
\Phi^{\bf\scriptscriptstyle R}_\alpha = \left( \Phi^{\bf\scriptscriptstyle R}_{+, \alpha}, \Phi^{\bf\scriptscriptstyle R}_{-, \alpha} \right)
\, .
\end{align} 
For the gluon, ${\bf R} = \bit{1}$,
\begin{align}
&
\Phi^{\bf\scriptscriptstyle 1}_{+, a > 0} = g_a \, , \\ 
&
\Phi^{\bf\scriptscriptstyle 1}_{-, - 2} = \psi \bar\psi_{\rm s} \, , \quad
\Phi^{\bf\scriptscriptstyle 1}_{-, - 1} = \phi \bar\psi_{\rm s} \bar\psi_{\rm s} \, , \quad
\Phi^{\bf\scriptscriptstyle 1}_{-, 0} = \bar\psi \bar\psi_{\rm s} \bar\psi_{\rm s} \bar\psi_{\rm s} \, , \quad
\Phi^{\bf\scriptscriptstyle 1}_{-, a > 0} =  \bar{g}_a \bar\psi_{\rm s} \bar\psi_{\rm s} \bar\psi_{\rm s} \bar\psi_{\rm s}
\, .
\end{align}
For the fermion,  ${\bf R} = \bit{4}$,
\begin{align}
&
\Phi^{\bf\scriptscriptstyle 4}_{+, 0} = \psi \, , \quad \Phi^{\bf\scriptscriptstyle 4}_{+, a > 0} = g_a \psi_{\rm s} \, , \\
&
\Phi^{\bf\scriptscriptstyle 4}_{-, - 1} = \phi \bar\psi_{\rm s} \, , \quad
\Phi^{\bf\scriptscriptstyle 4}_{-, 0} = \bar\psi \bar\psi_{\rm s}\bar\psi_{\rm s} \, , \quad
\Phi^{\bf\scriptscriptstyle 4}_{-, a > 0} = \bar{g}_a \bar\psi_{\rm s} \bar\psi_{\rm s}\bar\psi_{\rm s} \, ,
\end{align}
For the scalar, ${\bf R} = \bit{6}$,
\begin{align}
&
\Phi^{\bf\scriptscriptstyle 6}_{+, 0} = \psi \psi_{\rm s}  \, , \quad \Phi^{\bf\scriptscriptstyle 6}_{+, a > 0} = g_a \psi_{\rm s} \psi_{\rm s}  \, , \\
&
\Phi^{\bf\scriptscriptstyle 6}_{-, - 1} = \phi \, , \quad
\Phi^{\bf\scriptscriptstyle 6}_{-, 0} = \bar\psi \bar\psi_{\rm s} \, , \quad
\Phi^{\bf\scriptscriptstyle 6}_{-, a > 0} = \bar{g}_a \bar\psi_{\rm s}\bar\psi_{\rm s} \, .
\end{align}
The antigluon, ${\bf R} = \bit{\bar{1}}$ and antifermion ${\bf R} = \bit{\bar{4}}$ are obtained from the ones introduced above by dressing all particle symbols on the right
of their definitions with the bars and interchanging the subscripts designating helicities of corresponding parent excitations on the left, i.e.,  $\Phi_+ \leftrightarrow \Phi_-$. 
Notice that zero-helicity excitations were included along in a vector of negative helicity ones. As we will see later on, the latter have a smooth limit to the former and so they 
are naturally combined together.

\section{Independent heptagon components}
\label{SectionIndepHeptagonComp}

To start our analysis of the heptagon, we choose, by now conventional, parametrization of the momentum twistors $Z_i$ ($i = 1, \dots, 7$), which is recalled in Appendix 
\ref{TwistorsHeptagonSection} with the particular tessellation exhibited in Fig.\ \ref{Heptagon}. Each superpentagon operator $\mathbb{P}$ develops a finite-term expansion 
in Grassmann variable $\kappa$ assigned to each pentagon (along with bosonic variables $\tau$, $\sigma$ and $\varphi$),
\begin{align}
\mathbb{P}
= \mathcal{P} + \kappa_A \mathcal{P}^A + \ft{1}{2!} \kappa_A \kappa_B \mathcal{P}^{AB} 
+ \ft{1}{3!} \kappa_A \kappa_B \kappa_C \mathcal{P}^{ABC} + \ft{1}{4!} \kappa_A \kappa_B \kappa_C \kappa_D \mathcal{P}^{ABCD} 
\, ,
\end{align}
with its top/bottom components corresponding to the singlet ${\bf 1}/{\bf \bar{1}}$ transitions, the subleading ones from top and bottom to ${\bf 4}$ and ${\bf\bar{4}}$ of SU(4),
respectively, and finally the middle one to the antisymetric ${\bf 6}$ dimensional representation. Then, for the heptagon, which contains three overlapping pentagons (see 
Fig.\ \ref{Heptagon}), the NMHV component of the superWilson loop will have the Grassmann expansion
\begin{align}
\mathbb{W}_7^{\rm NMHV}
=
\sum^4_{{r_1, r_2, r_3 = 1 \atop r_1 + r_2 + r_3 = 4}}
\kappa_1^{r_1} \kappa_2^{r_2} \kappa_3^{r_3}
\, 
W^{[r_1, r_2, r_3]}_7
\, ,
\end{align}
where,  for brevity, we do not display SU(4) indices but rather only show their powers. Not all of the components $W^{[r_1, r_2, r_3]}_7$ are independent. Many of them are 
related by mean of supersymmetric Ward identities \cite{Elvang:2009wd}. For the case at hand, there are just fifteen \cite{Elvang:2009wd,Basso:2014hfa}. Their map to 
Grassmann components of the superloop expanded in terms of the fermionic variables $\chi^A_i$ of the momentum supertwistors $\mathcal{Z}_i = (Z_i, \chi^A_i)$ was
established in Ref.\ \cite{Basso:2014hfa} and reads for the case at hand
\begin{align}
\label{MapKappaToChi}
W^{[r_1, r_2, r_3]}_7
=
\left. \partial_{\chi_1}^{r_1} \left[ (2361) \partial_{\chi_1} + (2367) \partial_{\chi_7} \right]^{r_2} \partial_{\chi_4}^{r_3} \, 
\mathbb{W}_7 \right|_{\chi_i = 0}
\, ,
\end{align}
where $(ijkl) \equiv \varepsilon_{IJKL} Z_i^I Z_j^J Z_k^K Z_l^L$.

The rules to construct the flux-tube integrand were worked out in a series of papers alluded to in the Introduction. With the matrix part made available through a constructive 
technique of Ref.\ \cite{Belitsky:2016vyq}, one can find any Grassmann component of the superloop. Postponing details of the algebra involved to subsequent sections, we 
merely restrict ourselves with presenting explicit results for the contribution of parent excitations to the independent 
$W^{[r_1, r_2, r_3]}_7$'s,
\begin{align}
\label{Wr1r2r3}
W_7^{[r_1, r_2, r_3]} 
= 
\sum_{\alpha_1, \alpha_2} {\rm e}^{- t_{\alpha_1} \tau_1 - t_{\alpha_2} \tau_2 + i h_{\alpha_1} \varphi_1 + i h_{\alpha_2} \varphi_2} 
\int \frac{du \, dv}{(2 \pi)^2} {\rm e}^{2 i \sigma_1 u + 2 i \sigma_2 v} I^{\bf\scriptscriptstyle R_1 | R_2}(\alpha_1, u |, \alpha_2, v)
\, .
\end{align}
Here $t_{\alpha_i}$'s are the twists  and ${\rm\bf R}_i$ is the SU(4) charge of the parent excitations exchanged in the transitions. For generic values of $r_i$'s, there is no 
symmetry between positive $h_\alpha > 0$ and negative (including zero) $h_\alpha \leq 0$ helicity states. They are in fact different and are separated into independent 
terms below. Without further ado, we list the fifteen integrands of the independent components of the NMHV heptagon at leading order in coupling casting them in the form
$I^{\bf\scriptscriptstyle R_1 | R_2}_{{\rm sign}[h_{\alpha_1}] | {\rm sign}[h_{\alpha_2}]}$.
\begin{itemize}
\item $[4,0,0] \ 0 \to \Phi^{\bf\scriptscriptstyle 1}_{\alpha_1} (u) \to \Phi^{\bf\scriptscriptstyle 1}_{\alpha_2} (v) \to 0$:
\end{itemize}
\begin{align}
I^{\bf 1|1}_{+|+} (\alpha_1, u |, \alpha_2, v)
&
= i (-1)^{\alpha_1 + 1} 
\frac{
\Gamma (1 + \ft{\alpha_1}{2} + i u) \Gamma ( \ft{\alpha_2}{2} + i v) \Gamma (\ft{\alpha_1 - \alpha_2}{2} - i u - i v) \Gamma ( \ft{\alpha_1 + \alpha_2}{2} + i u + i v)
}{
(v + i \ft{\alpha_2}{2}) \Gamma (\alpha_1) \Gamma (\alpha_2) \Gamma (1 + \ft{\alpha_1 - \alpha_2}{2} + i u + i v)
}
\, , \\
I^{\bf 1|1}_{-|+} (\alpha_1, u |, \alpha_2, v)
&
= (-1)^{\alpha_1 + \alpha_2} 
\frac{
\Gamma (4 + \ft{\alpha_1}{2} + i u) \Gamma (\ft{\alpha_2}{2} + i v) \Gamma (1 + \ft{\alpha_1 + \alpha_2}{2} - i u - i v)
}{
(u + i \ft{\alpha_1}{2}) (v + i \ft{\alpha_2}{2}) \Gamma (4 + \alpha_1) \Gamma (\alpha_2)
}
\, , \\
I^{\bf 1|1}_{+|-} (\alpha_1, u |, \alpha_2, v)
&
= (-1)^{\alpha_1 + \alpha_2 + 1} 
\frac{
\Gamma (1 + \ft{\alpha_1}{2} + i u) \Gamma (1+ \ft{\alpha_2}{2} + i v) \Gamma (1 + \ft{\alpha_1 + \alpha_2}{2} - i u - i v)
}{
\Gamma (\alpha_1) \Gamma (4 + \alpha_2)
}
\, , \\
I^{\bf 1|1}_{-|-} (\alpha_1, u |, \alpha_2, v)
&
= i (-1)^{\alpha_1 + 1} 
\\
&\times\frac{
\Gamma (4 + \ft{\alpha_1}{2} + i u) \Gamma (1 + \ft{\alpha_2}{2} + i v) \Gamma (\ft{\alpha_1 - \alpha_2}{2} - i u - i v) \Gamma (4 + \ft{\alpha_1 + \alpha_2}{2} + i u + i v)
}{
(u + i \ft{\alpha_1}{2}) \Gamma (4 + \alpha_1) \Gamma (4 + \alpha_2) \Gamma (1 + \ft{\alpha_1 - \alpha_2}{2} + i u + i v)
}
\, . \nonumber
\end{align}

\begin{itemize}
\item $[3,0,1] \ 0 \to \Phi^{\bf\scriptscriptstyle 4}_{\alpha_1} (u) \to \Phi^{\bf\scriptscriptstyle 4}_{\alpha_2} (v) \to 0$:
\end{itemize}
\begin{align}
I^{\bf 4|4}_{+|+} (\alpha_1, u |, \alpha_2, v)
&
= i (-1)^{\alpha_1} 
\\
&\times
\frac{
\Gamma (1 + \ft{\alpha_1}{2} + i u) \Gamma (1 + \ft{\alpha_2}{2} + i v) \Gamma (\ft{\alpha_1 - \alpha_2}{2} - i u - i v) \Gamma (1 + \ft{\alpha_1 + \alpha_2}{2} + i u + i v)
}{
(v + i \ft{\alpha_2}{2}) \Gamma (1 + \alpha_1) \Gamma (1 + \alpha_2) \Gamma (1 + \ft{\alpha_1 - \alpha_2}{2} + i u + i v)
}
\, , \nonumber\\
I^{\bf 4|4}_{-|+} (\alpha_1, u |, \alpha_2, v)
&
= (-1)^{\alpha_1 + \alpha_2} 
\frac{
\Gamma (3 + \ft{\alpha_1}{2} + i u) \Gamma (1 + \ft{\alpha_2}{2} + i v) \Gamma (1 + \ft{\alpha_1 + \alpha_2}{2} - i u - i v)
}{
(u + i \ft{\alpha_1}{2}) (v + i \ft{\alpha_2}{2}) \Gamma (3 + \alpha_1) \Gamma (1 + \alpha_2)
}
\, , \\
I^{\bf 4|4}_{+|-} (\alpha_1, u |, \alpha_2, v)
&
= (-1)^{\alpha_1 + \alpha_2 + 1} 
\frac{
\Gamma (1 + \ft{\alpha_1}{2} + i u) \Gamma (1+ \ft{\alpha_2}{2} + i v) \Gamma (1 + \ft{\alpha_1 + \alpha_2}{2} - i u - i v)
}{
\Gamma (1 + \alpha_1) \Gamma (3 + \alpha_2)
}
\, , \nonumber\\
I^{\bf 4|4}_{-|-} (\alpha_1, u |, \alpha_2, v)
&
= i (-1)^{\alpha_1} 
\\
&
\times\frac{
\Gamma (3 + \ft{\alpha_1}{2} + i u) \Gamma (1 + \ft{\alpha_2}{2} + i v) \Gamma (\ft{\alpha_1 - \alpha_2}{2} - i u - i v) \Gamma (3 + \ft{\alpha_1 + \alpha_2}{2} + i u + i v)
}{
(u + i \ft{\alpha_1}{2}) \Gamma (3 + \alpha_1) \Gamma (3 + \alpha_2) \Gamma (1 + \ft{\alpha_1 - \alpha_2}{2} + i u + i v)
}
\, . \nonumber
\end{align}

\begin{itemize}
\item $[2,0,2] \ 0 \to \Phi^{\bf\scriptscriptstyle 6}_{\alpha_1} (u) \to \Phi^{\bf\scriptscriptstyle 6}_{\alpha_2} (v) \to 0$:
\end{itemize}
\begin{align}
I^{\bf 6|6}_{+|+} (\alpha_1, u |, \alpha_2, v)
&
= i (-1)^{\alpha_1 + 1} 
\\
&\times\frac{
\Gamma (1 + \ft{\alpha_1}{2} + i u) \Gamma (2 + \ft{\alpha_2}{2} + i v) \Gamma (\ft{\alpha_1 - \alpha_2}{2} - i u - i v) \Gamma (2 + \ft{\alpha_1 + \alpha_2}{2} + i u + i v)
}{
(v + i \ft{\alpha_2}{2}) \Gamma (2 + \alpha_1) \Gamma (2 + \alpha_2) \Gamma (1 + \ft{\alpha_1 - \alpha_2}{2} + i u + i v)
}
\, , \nonumber\\
I^{\bf 6|6}_{-|+} (\alpha_1, u |, \alpha_2, v)
&
= (-1)^{\alpha_1 + \alpha_2} 
\frac{
\Gamma (2 + \ft{\alpha_1}{2} + i u) \Gamma (2 + \ft{\alpha_2}{2} + i v) \Gamma (1 + \ft{\alpha_1 + \alpha_2}{2} - i u - i v)
}{
(u + i \ft{\alpha_1}{2}) (v + i \ft{\alpha_2}{2}) \Gamma (2 + \alpha_1) \Gamma (2 + \alpha_2)
}
\, , \\
I^{\bf 6|6}_{+|-} (\alpha_1, u |, \alpha_2, v)
&
= (-1)^{\alpha_1 + \alpha_2 + 1} 
\frac{
\Gamma (1 + \ft{\alpha_1}{2} + i u) \Gamma (1+ \ft{\alpha_2}{2} + i v) \Gamma (1 + \ft{\alpha_1 + \alpha_2}{2} - i u - i v)
}{
\Gamma (2 + \alpha_1) \Gamma (2 + \alpha_2)
}
\, , \nonumber\\
I^{\bf 6|6}_{-|-} (\alpha_1, u |, \alpha_2, v)
&
= i (-1)^{\alpha_1 + 1} 
\\
&\times\frac{
\Gamma (2 + \ft{\alpha_1}{2} + i u) \Gamma (1 + \ft{\alpha_2}{2} + i v) \Gamma (\ft{\alpha_1 - \alpha_2}{2} - i u - i v) \Gamma (2 + \ft{\alpha_1 + \alpha_2}{2} + i u + i v)
}{
(u + i \ft{\alpha_1}{2}) \Gamma (2 + \alpha_1) \Gamma (2 + \alpha_2) \Gamma (1 + \ft{\alpha_1 - \alpha_2}{2} + i u + i v)
}
\, . \nonumber
\end{align}

\begin{itemize}
\item $[2,2,0] \ 0 \to \Phi^{\bf\scriptscriptstyle 6}_{\alpha_1} (u) \to \Phi^{\bf\scriptscriptstyle 1}_{\alpha_2} (v) \to 0$:
\end{itemize}
\begin{align}
I^{\bf 6|1}_{+|+} (\alpha_1, u |, \alpha_2, v)
&
= i (-1)^{\alpha_1 + 1} 
\\
&\times
\frac{
\Gamma (1 + \ft{\alpha_1}{2} + i u) \Gamma ( \ft{\alpha_2}{2} + i v) \Gamma (\ft{\alpha_1 - \alpha_2}{2} - i u - i v) \Gamma (2 + \ft{\alpha_1 + \alpha_2}{2} + i u + i v)
}{
(v + i \ft{\alpha_2}{2}) \Gamma (2 + \alpha_1) \Gamma (\alpha_2) \Gamma (1 + \ft{\alpha_1 - \alpha_2}{2} + i u + i v)
}
\, , \nonumber\\
I^{\bf 6|1}_{-|+} (\alpha_1, u |, \alpha_2, v)
&
= (-1)^{\alpha_1 + \alpha_2} 
\frac{
\Gamma (2 + \ft{\alpha_1}{2} + i u) \Gamma (\ft{\alpha_2}{2} + i v) \Gamma (1 + \ft{\alpha_1 + \alpha_2}{2} - i u - i v)
}{
(u + i \ft{\alpha_1}{2}) (v + i \ft{\alpha_2}{2}) \Gamma (2 + \alpha_1) \Gamma  (\alpha_2)
}
\, , \\
I^{\bf 6|1}_{+|-} (\alpha_1, u |, \alpha_2, v)
&
= (-1)^{\alpha_1 + \alpha_2 + 1} 
(2 + \ft{\alpha_1 + \alpha_2}{2} + i u + i v ) (3 + \ft{\alpha_1 + \alpha_2}{2} + i u + i v )
\\
&\times
\frac{
\Gamma (1 + \ft{\alpha_1}{2} + i u) \Gamma (1+ \ft{\alpha_2}{2} + i v) \Gamma (1 + \ft{\alpha_1 + \alpha_2}{2} - i u - i v)
}{
\Gamma (2 + \alpha_1) \Gamma (4 + \alpha_2)
}
\, , \nonumber\\
I^{\bf 6|1}_{-|-} (\alpha_1, u |, \alpha_2, v)
&
= i (-1)^{\alpha_1 + 1} 
\\
&
\times
\frac{
\Gamma (2 + \ft{\alpha_1}{2} + i u) \Gamma (1 + \ft{\alpha_2}{2} + i v) \Gamma (\ft{\alpha_1 - \alpha_2}{2} - i u - i v) \Gamma (4 + \ft{\alpha_1 + \alpha_2}{2} + i u + i v)
}{
(u + i \ft{\alpha_1}{2}) \Gamma (2 + \alpha_1) \Gamma (4 + \alpha_2) \Gamma (1 + \ft{\alpha_1 - \alpha_2}{2} + i u + i v)
}
\, . \nonumber
\end{align}

\begin{itemize}
\item $[2,1,1] \ 0 \to \Phi^{\bf\scriptscriptstyle 6}_{\alpha_1} (u) \to \Phi^{\bf\scriptscriptstyle 4}_{\alpha_2} (v) \to 0$:
\end{itemize}
\begin{align}
I^{\bf 6|4}_{+|+} (\alpha_1, u |, \alpha_2, v)
&
= i (-1)^{\alpha_1} 
\\
&
\times
\frac{
\Gamma (1 + \ft{\alpha_1}{2} + i u) \Gamma (1 + \ft{\alpha_2}{2} + i v) \Gamma (\ft{\alpha_1 - \alpha_2}{2} - i u - i v) \Gamma (2 + \ft{\alpha_1 + \alpha_2}{2} + i u + i v)
}{
(v + i \ft{\alpha_2}{2}) \Gamma (2 + \alpha_1) \Gamma (1 + \alpha_2) \Gamma (1 + \ft{\alpha_1 - \alpha_2}{2} + i u + i v)
}
\, , \nonumber\\
I^{\bf 6|4}_{-|+} (\alpha_1, u |, \alpha_2, v)
&
= (-1)^{\alpha_1 + \alpha_2 + 1} 
\frac{
\Gamma (2 + \ft{\alpha_1}{2} + i u) \Gamma (1 + \ft{\alpha_2}{2} + i v) \Gamma (1 + \ft{\alpha_1 + \alpha_2}{2} - i u - i v)
}{
(u + i \ft{\alpha_1}{2}) (v + i \ft{\alpha_2}{2}) \Gamma (2 + \alpha_1) \Gamma (1 + \alpha_2)
}
\, , \\
I^{\bf 6|4}_{+|-} (\alpha_1, u |, \alpha_2, v)
&
= (-1)^{\alpha_1 + \alpha_2 + 1} 
\\
&\times
(2 + \ft{\alpha_1 + \alpha_2}{2} + i u + i v )
\frac{
\Gamma (1 + \ft{\alpha_1}{2} + i u) \Gamma (1+ \ft{\alpha_2}{2} + i v) \Gamma (1 + \ft{\alpha_1 + \alpha_2}{2} - i u - i v)
}{
\Gamma (2 + \alpha_1) \Gamma (3 + \alpha_2)
}
\, , \nonumber\\
I^{\bf 6|4}_{-|-} (\alpha_1, u |, \alpha_2, v)
&
= i (-1)^{\alpha_1 + 1} 
\\
&\frac{
\Gamma (2 + \ft{\alpha_1}{2} + i u) \Gamma (1 + \ft{\alpha_2}{2} + i v) \Gamma (\ft{\alpha_1 - \alpha_2}{2} - i u - i v) \Gamma (3 + \ft{\alpha_1 + \alpha_2}{2} + i u + i v)
}{
(u + i \ft{\alpha_1}{2}) \Gamma (2 + \alpha_1) \Gamma (3 + \alpha_2) \Gamma (1 + \ft{\alpha_1 - \alpha_2}{2} + i u + i v)
}
\, . \nonumber
\end{align}

\begin{itemize}
\item $[3,1,0] \ 0 \to \Phi^{\bf\scriptscriptstyle 4}_{\alpha_1} (u) \to \Phi^{\bf\scriptscriptstyle 1}_{\alpha_2} (v) \to 0$:
\end{itemize}
\begin{align}
I^{\bf 4|1}_{+|+} (\alpha_1, u |, \alpha_2, v)
&
= i (-1)^{\alpha_1 + 1} 
\\
&\times
\frac{
\Gamma (1 + \ft{\alpha_1}{2} + i u) \Gamma (\ft{\alpha_2}{2} + i v) \Gamma (\ft{\alpha_1 - \alpha_2}{2} - i u - i v) \Gamma (1 + \ft{\alpha_1 + \alpha_2}{2} + i u + i v)
}{
(v + i \ft{\alpha_2}{2}) \Gamma (1 + \alpha_1) \Gamma (\alpha_2) \Gamma (1 + \ft{\alpha_1 - \alpha_2}{2} + i u + i v)
}
\, , \nonumber\\
I^{\bf 4|1}_{-|+} (\alpha_1, u |, \alpha_2, v)
&
= (-1)^{\alpha_1 + \alpha_2 + 1} 
\frac{
\Gamma (3 + \ft{\alpha_1}{2} + i u) \Gamma (\ft{\alpha_2}{2} + i v) \Gamma (1 + \ft{\alpha_1 + \alpha_2}{2} - i u - i v)
}{
(u + i \ft{\alpha_1}{2}) (v + i \ft{\alpha_2}{2}) \Gamma (3 + \alpha_1) \Gamma (\alpha_2)
}
\, , \\
I^{\bf 4|1}_{+|-} (\alpha_1, u |, \alpha_2, v)
&
= (-1)^{\alpha_1 + \alpha_2 + 1} 
\\
&\times
(2 + \ft{\alpha_1 + \alpha_2}{2} + i u + i v )
\frac{
\Gamma (1 + \ft{\alpha_1}{2} + i u) \Gamma (1+ \ft{\alpha_2}{2} + i v) \Gamma (1 + \ft{\alpha_1 + \alpha_2}{2} - i u - i v)
}{
\Gamma (1 + \alpha_1) \Gamma (4 + \alpha_2)
}
\, , \nonumber\\
I^{\bf 4|1}_{-|-} (\alpha_1, u |, \alpha_2, v)
&
= i (-1)^{\alpha_1} 
\\
&\times\frac{
\Gamma (3 + \ft{\alpha_1}{2} + i u) \Gamma (1 + \ft{\alpha_2}{2} + i v) \Gamma (\ft{\alpha_1 - \alpha_2}{2} - i u - i v) \Gamma (4 + \ft{\alpha_1 + \alpha_2}{2} + i u + i v)
}{
(u + i \ft{\alpha_1}{2}) \Gamma (3 + \alpha_1) \Gamma (4 + \alpha_2) \Gamma (1 + \ft{\alpha_1 - \alpha_2}{2} + i u + i v)
}
\, . \nonumber
\end{align}

\begin{itemize}
\item $[0,4,0] \ 0 \to \Phi^{\bf\scriptscriptstyle \bar{1}}_{\alpha_1} (u) \to \Phi^{\bf\scriptscriptstyle 1}_{\alpha_2} (v) \to 0$:
\end{itemize}
\begin{align}
I^{\bf \bar{1}|1}_{+|+} (\alpha_1, u |, \alpha_2, v)
&
= i (-1)^{\alpha_1 + 1} 
\\
&\times\frac{
\Gamma (1 + \ft{\alpha_1}{2} + i u) \Gamma ( \ft{\alpha_2}{2} + i v) \Gamma (\ft{\alpha_1 - \alpha_2}{2} - i u - i v) \Gamma ( 4 + \ft{\alpha_1 + \alpha_2}{2} + i u + i v)
}{
(v + i \ft{\alpha_2}{2}) \Gamma (4 + \alpha_1) \Gamma (\alpha_2) \Gamma (1 + \ft{\alpha_1 - \alpha_2}{2} + i u + i v)
}
\, , \nonumber\\
I^{\bf \bar{1}|1}_{-|+} (\alpha_1, u |, \alpha_2, v)
&
= (-1)^{\alpha_1 + \alpha_2} 
\frac{
\Gamma ( \ft{\alpha_1}{2} + i u) \Gamma (\ft{\alpha_2}{2} + i v) \Gamma (1 + \ft{\alpha_1 + \alpha_2}{2} - i u - i v)
}{
(u + i \ft{\alpha_1}{2}) (v + i \ft{\alpha_2}{2}) \Gamma ( \alpha_1) \Gamma (\alpha_2)
}
\, , \\
I^{\bf \bar{1}|1}_{+|-} (\alpha_1, u |, \alpha_2, v)
&
= (-1)^{\alpha_1 + \alpha_2 + 1} 
(2 + \ft{\alpha_1 + \alpha_2}{2} + i u + i v ) (3 + \ft{\alpha_1 + \alpha_2}{2} + i u + i v ) (4 + \ft{\alpha_1 + \alpha_2}{2} + i u + i v )
\nonumber\\
& \times (5 + \ft{\alpha_1 + \alpha_2}{2} + i u + i v )
\frac{
\Gamma (1 + \ft{\alpha_1}{2} + i u) \Gamma (1+ \ft{\alpha_2}{2} + i v) \Gamma (1 + \ft{\alpha_1 + \alpha_2}{2} - i u - i v)
}{
\Gamma (4 + \alpha_1) \Gamma (4 + \alpha_2)
}
\, , \\
I^{\bf \bar{1}|1}_{-|-} (\alpha_1, u |, \alpha_2, v)
&
= i (-1)^{\alpha_1 + 1} 
\\
&\times
\frac{
\Gamma (\ft{\alpha_1}{2} + i u) \Gamma (1 + \ft{\alpha_2}{2} + i v) \Gamma (\ft{\alpha_1 - \alpha_2}{2} - i u - i v) \Gamma (4 + \ft{\alpha_1 + \alpha_2}{2} + i u + i v)
}{
(u + i \ft{\alpha_1}{2}) \Gamma (\alpha_1) \Gamma (4 + \alpha_2) \Gamma (1 + \ft{\alpha_1 - \alpha_2}{2} + i u + i v)
}
\, . \nonumber
\end{align}

\begin{itemize}
\item $[0,3,1] \ 0 \to \Phi^{\bf\scriptscriptstyle \bar{1}}_{\alpha_1} (u) \to \Phi^{\bf\scriptscriptstyle 4}_{\alpha_2} (v) \to 0$:
\end{itemize}
\begin{align}
I^{\bf \bar{1}|4}_{+|+} (\alpha_1, u |, \alpha_2, v)
&
= i (-1)^{\alpha_1} 
\\
&\times\frac{
\Gamma (1 + \ft{\alpha_1}{2} + i u) \Gamma (1 + \ft{\alpha_2}{2} + i v) \Gamma (\ft{\alpha_1 - \alpha_2}{2} - i u - i v) \Gamma ( 4 + \ft{\alpha_1 + \alpha_2}{2} + i u + i v)
}{
(v + i \ft{\alpha_2}{2}) \Gamma (4 + \alpha_1) \Gamma (1 + \alpha_2) \Gamma (1 + \ft{\alpha_1 - \alpha_2}{2} + i u + i v)
}
\, , \nonumber\\
I^{\bf \bar{1}|4}_{-|+} (\alpha_1, u |, \alpha_2, v)
&
= (-1)^{\alpha_1 + \alpha_2 + 1} 
\frac{
\Gamma ( \ft{\alpha_1}{2} + i u) \Gamma (1 + \ft{\alpha_2}{2} + i v) \Gamma (1 + \ft{\alpha_1 + \alpha_2}{2} - i u - i v)
}{
(u + i \ft{\alpha_1}{2}) (v + i \ft{\alpha_2}{2}) \Gamma ( \alpha_1) \Gamma (1 + \alpha_2)
}
\, , \\
I^{\bf \bar{1}|4}_{+|-} (\alpha_1, u |, \alpha_2, v)
&
= (-1)^{\alpha_1 + \alpha_2 + 1} 
(2 + \ft{\alpha_1 + \alpha_2}{2} + i u + i v ) (3 + \ft{\alpha_1 + \alpha_2}{2} + i u + i v ) (4 + \ft{\alpha_1 + \alpha_2}{2} + i u + i v )
\nonumber\\
& \times 
\frac{
\Gamma (1 + \ft{\alpha_1}{2} + i u) \Gamma (1+ \ft{\alpha_2}{2} + i v) \Gamma (1 + \ft{\alpha_1 + \alpha_2}{2} - i u - i v)
}{
\Gamma (4 + \alpha_1) \Gamma (3 + \alpha_2)
}
\, , \\
I^{\bf \bar{1}|4}_{-|-} (\alpha_1, u |, \alpha_2, v)
&
= i (-1)^{\alpha_1 + 1} 
\\
&\times
\frac{
\Gamma (\ft{\alpha_1}{2} + i u) \Gamma (1 + \ft{\alpha_2}{2} + i v) \Gamma (\ft{\alpha_1 - \alpha_2}{2} - i u - i v) \Gamma (3 + \ft{\alpha_1 + \alpha_2}{2} + i u + i v)
}{
(u + i \ft{\alpha_1}{2}) \Gamma (\alpha_1) \Gamma (3 + \alpha_2) \Gamma (1 + \ft{\alpha_1 - \alpha_2}{2} + i u + i v)
}
\, . \nonumber
\end{align}

\begin{itemize}
\item $[1,2,1] \ 0 \to \Phi^{\bf\scriptscriptstyle \bar{4}}_{\alpha_1} (u) \to \Phi^{\bf\scriptscriptstyle 4}_{\alpha_2} (v) \to 0$:
\end{itemize}
\begin{align}
I^{\bf \bar{4}|4}_{+|+} (\alpha_1, u |, \alpha_2, v)
&
= i (-1)^{\alpha_1} 
\\
&\times\frac{
\Gamma (1 + \ft{\alpha_1}{2} + i u) \Gamma (1 + \ft{\alpha_2}{2} + i v) \Gamma (\ft{\alpha_1 - \alpha_2}{2} - i u - i v) \Gamma ( 3 + \ft{\alpha_1 + \alpha_2}{2} + i u + i v)
}{
(v + i \ft{\alpha_2}{2}) \Gamma (3 + \alpha_1) \Gamma (1 + \alpha_2) \Gamma (1 + \ft{\alpha_1 - \alpha_2}{2} + i u + i v)
}
\, , \nonumber\\
I^{\bf \bar{4}|4}_{-|+} (\alpha_1, u |, \alpha_2, v)
&
= (-1)^{\alpha_1 + \alpha_2} 
\frac{
\Gamma ( 1 + \ft{\alpha_1}{2} + i u) \Gamma (1 + \ft{\alpha_2}{2} + i v) \Gamma (1 + \ft{\alpha_1 + \alpha_2}{2} - i u - i v)
}{
(u + i \ft{\alpha_1}{2}) (v + i \ft{\alpha_2}{2}) \Gamma ( 1 + \alpha_1) \Gamma (1 + \alpha_2)
}
\, , \\
I^{\bf \bar{4}|4}_{+|-} (\alpha_1, u |, \alpha_2, v)
&
= (-1)^{\alpha_1 + \alpha_2 + 1} 
(2 + \ft{\alpha_1 + \alpha_2}{2} + i u + i v ) (3 + \ft{\alpha_1 + \alpha_2}{2} + i u + i v ) 
\nonumber\\
& \times 
\frac{
\Gamma (1 + \ft{\alpha_1}{2} + i u) \Gamma (1+ \ft{\alpha_2}{2} + i v) \Gamma (1 + \ft{\alpha_1 + \alpha_2}{2} - i u - i v)
}{
\Gamma (3 + \alpha_1) \Gamma (3 + \alpha_2)
}
\, , \\
I^{\bf \bar{4}|4}_{-|-} (\alpha_1, u |, \alpha_2, v)
&
= i (-1)^{\alpha_1} 
\\
&\times
\frac{
\Gamma (1 + \ft{\alpha_1}{2} + i u) \Gamma (1 + \ft{\alpha_2}{2} + i v) \Gamma (\ft{\alpha_1 - \alpha_2}{2} - i u - i v) \Gamma (3 + \ft{\alpha_1 + \alpha_2}{2} + i u + i v)
}{
(u + i \ft{\alpha_1}{2}) \Gamma (1 + \alpha_1) \Gamma (3 + \alpha_2) \Gamma (1 + \ft{\alpha_1 - \alpha_2}{2} + i u + i v)
}
\, . \nonumber
\end{align}

The remaining 6 can be found from these by interchanging the top and bottom excitations. Namely,
\begin{itemize}
\item $[0,0,4] \ 0 \to \Phi^{\bf\scriptscriptstyle \bar{1}}_{\alpha_1} (u) \to \Phi^{\bf\scriptscriptstyle \bar{1}}_{\alpha_2} (v) \to 0$:
\end{itemize}
\begin{align}
I^{\bf \bar{1}|\bar{1}}_{+|+} (\alpha_1, u |, \alpha_2, v) 
&
= 
I^{\bf 1| 1}_{-|-} (\alpha_2, v |, \alpha_1, u)
\, , \\
I^{\bf \bar{1}|\bar{1}}_{-|+} (\alpha_1, u |, \alpha_2, v)
&
= 
I^{\bf 1| 1}_{-|+} (\alpha_2, v |, \alpha_1, u)
\, , \\
I^{\bf \bar{1}|\bar{1}}_{+|-} (\alpha_1, u |, \alpha_2, v)
&
= 
I^{\bf 1| 1}_{+|-} (\alpha_2, v |, \alpha_1, u)
\, , \\
I^{\bf \bar{1}|\bar{1}}_{-|-} (\alpha_1, u |, \alpha_2, v)
&
= 
I^{\bf 1| 1}_{+|+} (\alpha_2, v |, \alpha_1, u)
\, .
\end{align}

\begin{itemize}
\item $[1,0,3] \ 0 \to \Phi^{\bf\scriptscriptstyle \bar{4}}_{\alpha_1} (u) \to \Phi^{\bf\scriptscriptstyle \bar{4}}_{\alpha_2} (v) \to 0$:
\end{itemize}
\begin{align}
I^{\bf \bar{4}|\bar{4}}_{+|+} (\alpha_1, u |, \alpha_2, v) 
&
= 
I^{\bf 4| 4}_{-|-} (\alpha_2, v |, \alpha_1, u)
\, , \\
I^{\bf \bar{4}|\bar{4}}_{-|+} (\alpha_1, u |, \alpha_2, v)
&
= 
I^{\bf 4| 4}_{-|+} (\alpha_2, v |, \alpha_1, u)
\, , \\
I^{\bf \bar{4}|\bar{4}}_{+|-} (\alpha_1, u |, \alpha_2, v)
&
= 
I^{\bf 4| 4}_{+|-} (\alpha_2, v |, \alpha_1, u)
\, , \\
I^{\bf \bar{4}|\bar{4}}_{-|-} (\alpha_1, u |, \alpha_2, v)
&
= 
I^{\bf 4| 4}_{+|+} (\alpha_2, v |, \alpha_1, u)
\, .
\end{align}

\begin{itemize}
\item $[0,2,2] \ 0 \to \Phi^{\bf\scriptscriptstyle \bar{1}}_{\alpha_1} (u) \to \Phi^{\bf\scriptscriptstyle 6}_{\alpha_2} (v) \to 0$:
\end{itemize}
\begin{align}
I^{\bf \bar{1}| 6}_{+|+} (\alpha_1, u |, \alpha_2, v) 
&
= 
I^{\bf 6| 1}_{-|-} (\alpha_2, v |, \alpha_1, u)
\, , \\
I^{\bf \bar{1}| 6}_{-|+} (\alpha_1, u |, \alpha_2, v)
&
= 
I^{\bf 6| 1}_{-|+} (\alpha_2, v |, \alpha_1, u)
\, , \\
I^{\bf \bar{1}| 6}_{+|-} (\alpha_1, u |, \alpha_2, v)
&
=
I^{\bf 6| 1}_{+|-} (\alpha_2, v |, \alpha_1, u)
\, , \\
I^{\bf \bar{1}| 6}_{-|-} (\alpha_1, u |, \alpha_2, v)
&
= 
I^{\bf 6| 1}_{+|+} (\alpha_2, v |, \alpha_1, u)
\, .
\end{align}

\begin{itemize}
\item $[1,1,2] \ 0 \to \Phi^{\bf\scriptscriptstyle \bar{4}}_{\alpha_1} (u) \to \Phi^{\bf\scriptscriptstyle 6}_{\alpha_2} (v) \to 0$:
\end{itemize}
\begin{align}
I^{\bf \bar{4}| 6}_{+|+} (\alpha_1, u |, \alpha_2, v) 
&
= 
I^{\bf 6| 4}_{-|-} (\alpha_2, v |, \alpha_1, u)
\, , \\
I^{\bf \bar{4}| 6}_{-|+} (\alpha_1, u |, \alpha_2, v)
&
= 
I^{\bf 6| 4}_{-|+} (\alpha_2, v |, \alpha_1, u)
\, , \\
I^{\bf \bar{4}| 6}_{+|-} (\alpha_1, u |, \alpha_2, v)
&
=
I^{\bf 6| 4}_{+|-} (\alpha_2, v |, \alpha_1, u)
\, , \\
I^{\bf \bar{4}| 6}_{-|-} (\alpha_1, u |, \alpha_2, v)
&
= 
I^{\bf 6| 4}_{+|+} (\alpha_2, v |, \alpha_1, u)
\, .
\end{align}

\begin{itemize}
\item $[0,1,3] \ 0 \to \Phi^{\bf\scriptscriptstyle \bar{1}}_{\alpha_1} (u) \to \Phi^{\bf\scriptscriptstyle \bar{4}}_{\alpha_2} (v) \to 0$:
\end{itemize}
\begin{align}
I^{\bf \bar{1}| \bar{4}}_{+|+} (\alpha_1, u |, \alpha_2, v) 
&
= 
I^{\bf 4| 1}_{-|-} (\alpha_2, v |, \alpha_1, u)
\, , \\
I^{\bf \bar{1}| \bar{4}}_{-|+} (\alpha_1, u |, \alpha_2, v)
&
= 
I^{\bf 4| 1}_{-|+} (\alpha_2, v |, \alpha_1, u)
\, , \\
I^{\bf \bar{1}| \bar{4}}_{+|-} (\alpha_1, u |, \alpha_2, v)
&
=
I^{\bf 4| 1}_{+|-} (\alpha_2, v |, \alpha_1, u)
\, , \\
I^{\bf \bar{1}| \bar{4}}_{-|-} (\alpha_1, u |, \alpha_2, v)
&
= 
I^{\bf 4| 1}_{+|+} (\alpha_2, v |, \alpha_1, u)
\, .
\end{align}

\begin{itemize}
\item $[1,3,0] \ 0 \to \Phi^{\bf\scriptscriptstyle \bar{4}}_{\alpha_1} (u) \to \Phi^{\bf\scriptscriptstyle 1}_{\alpha_2} (v) \to 0$:
\end{itemize}
\begin{align}
I^{\bf \bar{4}| 1}_{+|+} (\alpha_1, u |, \alpha_2, v) 
&
= 
I^{\bf \bar{1}| 4}_{-|-} (\alpha_2, v |, \alpha_1, u)
\, , \\
I^{\bf \bar{4}| 1}_{-|+} (\alpha_1, u |, \alpha_2, v)
&
= 
I^{\bf \bar{1}| 4}_{-|+} (\alpha_2, v |, \alpha_1, u)
\, , \\
I^{\bf \bar{1}| 4}_{+|-} (\alpha_1, u |, \alpha_2, v)
&
=
I^{\bf 4| 1}_{+|-} (\alpha_2, v |, \alpha_1, u)
\, , \\
I^{\bf \bar{4}| 1}_{-|-} (\alpha_1, u |, \alpha_2, v)
&
= 
I^{\bf \bar{1}| 4}_{+|+} (\alpha_2, v |, \alpha_1, u)
\, .
\end{align}
Using the map \re{MapKappaToChi}, it is straightforward to cross-check the correctness of the above expressions against explicit heptagon data \cite{Bourjaily:2013mma}.

Having found the integrands for parent excitations, we are in a position to construct these for effective particles by dressing the former with infinite cloud of
small fermion-antifermions pairs.

\section{Warm-up: resummation of pairs for hexagon}

To start with the resummation of small fermion-antifermion pairs, let us recall the structure of the result for the NMHV hexagon studied in Ref.\ \cite{Cordova:2016woh}.
It will be sufficient to demonstrate it for a specific component, which we choose to be $W_6^{[2,2]}$, ---in the notation analogous to the one adopted for the heptagon,--- 
\begin{align}
W_6^{[2,2]}
=
-\frac{e^{\tau -\sigma }}{e^{2 \tau }+1}
+
\frac{e^{-\sigma -\tau }}{2 e^{\sigma -\tau } \cos \varphi+e^{2 \sigma } + e^{-2 \tau }+1}
\, ,
\end{align}
and, in fact, for just one Fourier component in the angle $\varphi$,
\begin{align}
\vev{W_6^{[2,2]}}_0
&\equiv
\frac{1}{4 \pi} \int_0^{4 \pi} d \varphi \, W_6^{[2,2]}
\nonumber\\
&=
-\frac{e^{\tau -\sigma }}{e^{2 \tau }+1}
+
\frac{e^{-\sigma -\tau }}{\sqrt{(1 + e^{2 \sigma } + e^{-2 \tau })^2 - 4 e^{2 \sigma -2 \tau }}}
\, .
\end{align}
Its leading term in the collinear expansion as $\tau \to \infty$ corresponds to the contributions due to the exchange of helicity zero $\phi$-excitation with 
the flux-tube integrand
\begin{align}
I^{\bf 6} (u,0|v,0) =  \Gamma (-\ft12 - i u) \Gamma (\ft32 + i u)
\, .
\end{align}
To account for the entire infinite series of subleading terms in ${\rm e}^{- ( 2 n + 1) \tau}$, one has to calculate integrands of $0 \to \phi (\bar\psi_{\rm s} \psi_{\rm s})^n \to 0$ 
transitions for any $n$. Though an integral representation for the traced tensor part of pentagon transitions, dubbed as the matrix part, is know for the hexagon in term 
of an integral representation \cite{Basso:2015uxa}, practically, they can be evaluated for large but finite number of excitations only. Particular patterns of string formation 
involving small (anti)fermions with a parent excitation were established in Ref.\ \cite{Cordova:2016woh} and used to conjecture a generic functional dependence on $n$,
\begin{align}
\vev{W_6^{[2,2]}}_0
= {\rm e}^{- \tau} \sum_{n = 0}^\infty \frac{(- {\rm e}^{- 2 \tau})^n}{(n!)^2} \int \frac{d u}{2 \pi}
{\rm e}^{2 i u \sigma} \frac{\Gamma (-\ft12 - i u) \Gamma^2 (\ft32 + n + i u)}{\Gamma (3/2 + i u)}
\, .
\end{align}
The summation of the series yields the result
\begin{align}
\vev{W_6^{[2,2]}}_0
&= {\rm e}^{- \tau} 
\int \frac{d u}{2 \pi}
{\rm e}^{2 i u \sigma} I^{\bf 6} (u,0|v,0) \, 
{_2 F_1}
\left.\left(
{ \ft32 + i u, \ft32 + i u \atop 1}
\right| - {\rm e}^{- 2 \tau}
\right)
\, ,
\end{align}
where the flux-tube $0 \to \phi \to 0$ integrand $I^{\bf 6}$ is accompanied by the single-channel sl(2) conformal block expressed in terms of the hypergeometric function 
${_2 F_1}$, as was observed earlier in \cite{Gaiotto:2011dt,{Cordova:2016woh}}. We review its construction in Appendix \ref{ConformalBlocksSection}. Below, when we 
will turn to the heptagon and adopt a partially reversed ideology and use explicit expression of conformal blocks as a guiding princile for infinite series 
resummation.

\section{Resummation of pairs for heptagon: an example}
\label{HeptagonExample}

Since the matrix part is available only via a constructive procedure, which has to be performed for every new set of excitations, analyses were done for a relatively small 
number of excitations. They provided the first few terms in the infinite series expansion, whose dependence on the orders in the expansion had to be guessed. To have
a proper guidance in this endeavor and unravel the general pattern, the unbroken sl(2) algebra of the leading-order flux-tube dynamics was used as a guiding principle to 
fix the form of the conformal blocks. In Appendix \ref{TwistorsHeptagonSection} we do this for the case without helicity weights of NMHV amplitudes, whose sole effect is 
to shift some numerical indices by constants in the fifteen independent NMHV components.

%%%%%%%%%%%%%%%%%%%%%%%%%%%%%%%%%%%%%%%%%%%%%%%%%%%%%%%%%%%%%%%%%%%%%
%            Figure
%%%%%%%%%%%%%%%%%%%%%%%%%%%%%%%%%%%%%%%%%%%%%%%%%%%%%%%%%%%%%%%%%%%%%
\begin{figure}[t]
\begin{center}
\mbox{
\begin{picture}(0,300)(200,0)
\put(0,-20){\insertfig{15}{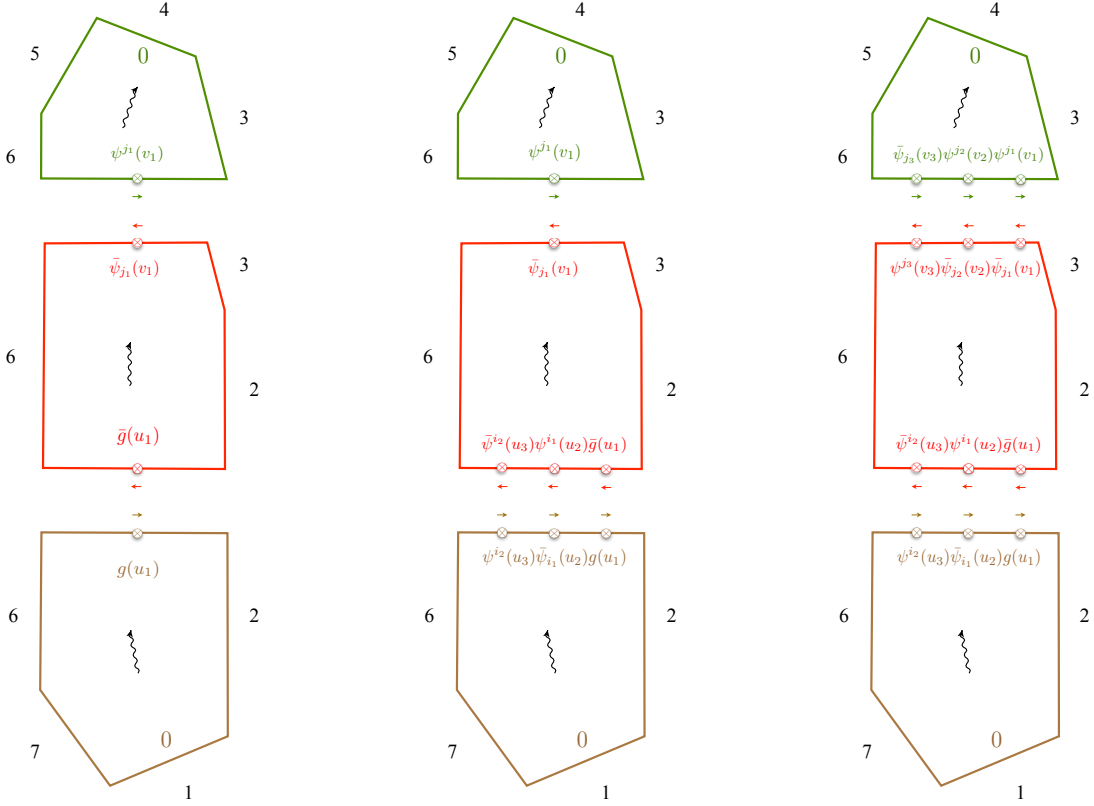}}
\end{picture}
}
\end{center}
\caption{ \label{DeconstructionPic} Deconstruction of the $0 \to \bar{g} (u_1) (\psi \bar\psi)^{n_1} \to  \psi (u_2)  (\psi \bar\psi)^{n_2} \to 0$ heptagon in terms of 
pentagon transitions with increasing number of flux-tube excitations with $(n_1, n_2)$ being $(0,0)$, (1,0) and $(1,1)$ in the left, central and right panels,
respectively.}
\end{figure}
%%%%%%%%%%%%%%%%%%%%%%%%%%%%%%%%%%%%%%%%%%%%%%%%%%%%%%%%%%%%%%%%%%%%%

Let us demonstrate this technology for a specific transition, say, with a charged middle pentagon $W^{[0,3,1]}_{7}$. The first few terms in the pentagon operator
product expansion read
\begin{align}
W^{[0,3,1]}_{7} 
= 
{\rm e}^{- i \varphi_1 + i \varphi_2/2} 
\big[
&
{\rm e}^{- \tau_1 - \tau_2} W^{[0,3,1]}_{7 (0 \to \bar{g} \to \psi \to 0)}
\\
+ 
&
{\rm e}^{- 3 \tau_1 - \tau_2} W^{[0,3,1]}_{7 (0 \to \bar{g} (\psi_{\rm s} \bar{\psi}_{\rm s}) \to \psi \to 0)}
+ {\rm e}^{- 3 \tau_1 - 3 \tau_2} W^{[0,3,1]}_{7 (0 \to \bar{g} (\psi_{\rm s} \bar{\psi}_{\rm s}) \to \psi  (\psi_{\rm s} \bar{\psi}_{\rm s}) \to 0)}
+ \dots
\big]
\, , \nonumber
\end{align}
where we exhibited the explicit transitions as labels and their deconstruction in terms of pentagon form factors is shown in Fig.\ \ref{DeconstructionPic} for each
term, respectively.

The leading term $0 \to \bar{g} \to \psi \to 0$, shown in the leftmost panel of Fig.\ \ref{DeconstructionPic}, does not involve any complicated tensor structures and 
reads
\begin{align}
\label{OneToOne}
W^{[0,3,1]}_{7 (0 \to \bar{g} \to \psi \to 0)}
&=
\int
d \widehat\mu_g (u_1) d\widehat\mu_\psi (v_1) \sqrt{x^+[u_1] x^-[u_2]} x[v_1] P_{\bar{g}|\psi} (- u_1 | v_1)
\\
&
=
\int \frac{d u_1 \, d v_1}{(2 \pi)^2}
{\rm e}^{2 i u_1 \sigma_1 + 2 i v_1 \sigma_2} 
\Gamma ( - \ft12 + i u_1) \Gamma (i v_1) \Gamma (\ft32 - i u_1 - i v_1)
\, . \nonumber
\end{align}
where here and below we use the convention
\begin{align}
d \widehat{\mu}_{\rm p} (u) = \frac{d u}{2 \pi} {\rm e}^{i \sigma p_{\rm p} (u)} \mu_{\rm p} (p)
\, ,
\end{align}
with $p_{\rm p} (u) = 2 u + O (g^2)$ for (anti)gluon, scalar and large anti(fermion) flux-tube excitations.

The next case in complexity is the $0 \to \bar{g} ( \psi_{\rm s} \bar\psi_{\rm s} ) \to \psi \to 0$ transition in the middle panel of Fig.\ \ref{DeconstructionPic}.

\begin{align}
\label{GbarToPsi3to1}
&
W^{[0,3,1]}_{7 (0 \to \bar{g} (\psi_{\rm s} \bar{\psi}_{\rm s}) \to \psi\to 0)}
=
\int
d \widehat\mu_g (u_1) d \widehat\mu_\psi (u_2) d \widehat\mu_\psi (u_3) d \widehat\mu_\psi (v_1) 
\\
&\qquad\qquad
\times\!
\sqrt{x^+[u_1] x^-[u_1]}  x[v_1] \frac{x[u_3]}{x[u_2]}
\Pi^{\bf 1}_{0|\bar\psi_{i_1} \psi^{i_2} } (0| u_2, u_3) [\Pi^{\bf 4}_{\bar\psi_{i_2} \psi^{i_1} | \bar\psi_{j_1}}]^{j_2} (- u_3, - u_2 | v_1)
[\Pi^{\bf\bar{4}}_{\psi^{j_1} |0}]_{j_2} (-v_1 |0)
\nonumber\\
&\qquad\qquad
\times
P_{0|\bar{g}\psi\bar\psi} (0| u_1, u_2, u_3) P_{\bar\psi\psi \bar{g} |\psi} (- u_3, - u_2, - u_1 |v_1)
P_{\psi |0} (-v_1|0)
\, . \nonumber
\end{align}
The two-to-one transition tensor involved in the decomposition is given in Eq.\ \re{2to1fermiontensor}, while  $[\Pi^{\bf\bar{4}}_{\psi^{j_1} |0}]_{j_2} (-v_1 |0)
= \delta_{j_2}^{j_1}$. Making use of the factorization property \re{FactorizedPentagon} of the multiparticle pentagons, we find, passing to the small fermion
sheet in variables $u_2$ and $u_3$,
\begin{align}
W^{[0,3,1]}_{7 (0 \to \bar{g} (\psi_{\rm s} \bar{\psi}_{\rm s}) \to \psi\to 0)}
&
=
\int
d \widehat\mu_g (u_1) d \widehat\mu_\psi (v_1) x[v_1] \sqrt{x^+ [u_1] x^- [u_1]} P_{\bar{g}|\psi} (- u_1 | v_1)
\\
&\times
\int_{C_-}  \frac{d u_2 \, d u_3 }{(2 \pi)^2}
\frac{ (u_2|u_3)_{-2} - 4 (u_2 | - v_1)_{-1}}{(u_2 | u_3)_2 (u_2|u_3)_{-2}}
\frac{u_3  }{(u_1|u_3)_{1/2} (u_1|u_3)_{-1/2} }
\, , \nonumber
\end{align}
where the integration contour $C_-$ runs clockwise over a half-moon contour in the lower half plane of the complex plane of the corresponding small-(anti)fermion 
rapidities. Evaluating these integrals by residues, we get
\begin{align}
&
W^{[0,3,1]}_{7 (0 \to \bar{g} (\psi_{\rm s} \bar{\psi}_{\rm s}) \to \psi\to 0)}
\\
&\qquad\qquad
=
-
\int
d \widehat\mu_g (u_1) d \widehat\mu_\psi (v_1) x[v_1] \sqrt{x^+ [u_1] x^- [u_1]} P_{\bar{g}|\psi} (- u_1 | v_1)
\left[
(u_1 - \ft{i}{2}) (u_1 + v_1 - \ft{5 i}{2})
\right]
\, , \nonumber
\end{align}
which differs from the one-to-one transition \re{OneToOne} by a second order polynomial in rapidities $u_1$ and $v_1$.

Finally, let us turn to $0 \to \bar{g} (\psi \bar\psi) \to \psi ( \psi \bar\psi ) \to 0$, which is shown in the right panel of Fig.\ \ref{DeconstructionPic}.
\begin{align}
\label{GbarToPsi3to3}
&
W^{[0,3,1]}_{7 (0 \to \bar{g} (\psi_{\rm s} \bar{\psi}_{\rm s}) \to \psi  (\psi_{\rm s} \bar{\psi}_{\rm s}) \to 0)}
=
\int
d \widehat\mu_g (u_1) d \widehat\mu_\psi (u_2) d \widehat\mu_\psi (u_3) d \widehat\mu_\psi (v_1) d \widehat\mu_\psi (v_2) d \widehat\mu_\psi (v_3) 
\\
&\qquad\qquad
\times
\sqrt{x^+[u_1] x^-[u_1]}  x[v_1] \frac{x[u_3]}{x[u_2]} \frac{x[v_2]}{x[v_3]} 
\nonumber\\
&
\times
\Pi^{\bf 1}_{0|\bar\psi_{i_1} \psi^{i_2} } (0| u_2, u_3) [\Pi^{\bf 4}_{\bar\psi_{i_2} \psi^{i_1} | \psi^{j_3} \bar\psi_{j_2} \bar\psi_{j_1}}]^{i_3} (- u_3, - u_2 | v_3, v_2, v_1)
[\Pi^{\bf\bar{4}}_{\psi^{j_1} \psi^{j_2} \bar\psi_{j_3} |0}]_{i_3} (-v_1, -v_2, -v_3|0)
\nonumber\\
&\qquad\qquad
\times
P_{0|\bar{g}\psi\bar\psi} (0| u_1, u_2, u_3) P_{\bar\psi\psi \bar{g} | \bar\psi \psi \psi} (- u_3, - u_2, - u_1 | v_3, v_2, v_1)
P_{\psi \psi \bar\psi |0} (-v_1, -v_2, -v_3|0)
\, . \nonumber
\end{align}
The dynamical part reads at leading order in coupling
\begin{align}
P_{0|\bar{g} \psi_{\rm s} \bar\psi_{\rm s}} (0 | u_1, u_2, u_3)
&
P_{\bar\psi_{\rm s} \psi_{\rm s} \bar{g} | \bar\psi_{\rm s} \psi_{\rm s} \psi} (- u_3, - u_2, - u_1 | v_3, v_2, v_1)
P_{\psi \psi_{\rm s} \bar\psi_{\rm s} |0} (- v_1, - v_2, - v_3 | 0)
\nonumber\\
&=
\frac{u_3^2 v_2^2}{u_2 v_3}
\frac{(u_1 |- v_3)_{-1/2} P_{\bar{g} | \psi} (- u_1 | v_1)}{(v_3 |-u_3)_0 (v_2 |- u_2)_0 (u_3| u_1)_{1/2}  (u_3| u_1)_{- 1/2}}
\, ,
\end{align}
after passing with (anti)fermions to the small sheet and making use of the leading order expressions for the pentagons. Evaluating the residues over the small 
fermions rapidities in the order $v_2 \to v_1 \to w_1 \to w_2$ of string formation, we find
\begin{align}
\res\limits_{v_3 = v_1 - 2 i}
\res\limits_{v_2 = v_1 - i}
\res\limits_{u_2 = u_1 - 5i/2}
\res\limits_{u_3 = u_1 - i/2}
&
\left[
\frac{u_3 v_2 (u_1 |- v_3)_{-1/2} 
\Pi^{\bf 1}_{0| \bar\psi_{i_1} \psi^{i_2} } [\Pi^{\bf 4}_{\bar\psi_{i_2} \psi^{i_1} | \psi^{j_3} \bar\psi_{j_2} \bar\psi_{j_1}}]^{i_3}
[\Pi^{\bf\bar{4}}_{\psi^{j_1} \psi^{j_2} \bar\psi_{j_3} |0}]_{i_3}
}{(v_3 |-u_3)_0 (v_2 |- u_2)_0 (u_3| u_1)_{1/2}  (u_3| u_1)_{- 1/2}}
\right]
\nonumber\\
&
=
(u_1 - \ft{i}{2}) (u_2 - i) (u_1 + u_2 - \ft{5 i}{2})  (u_1 + u_2 - \ft{7 i}{2})
\, .
\end{align}

Anticipating the emergence of the Appell function as a conformal block for amplitudes that effectively resums the infinite series in fermion-antifermion pairs,
we can immediately fix the indices of the former to the first few terms found above. We deduce
\begin{align}
\vev{W^{[0,3,1]}_{7}}_{1,-1/2}
&\equiv
\frac{1}{(4 \pi)^2} \int_0^{4 \pi} d \varphi_1 d \varphi_2 \, 
{\rm e}^{i \varphi_1 - i \varphi_2/2} 
W^{[0,3,1]}_{7} 
\\
&
= 
{\rm e}^{- \tau_1 - \tau_2}
\int \frac{d u \, d v}{(2 \pi)^2}
{\rm e}^{2 i u_1 \sigma_1 + 2 i u_2 \sigma_2} 
\Gamma ( - \ft12 + i u) \Gamma ( i v) \Gamma ( \ft32 - i u - i v)
\nonumber\\
&\qquad\qquad\qquad
\times F_2
\left. \left( { \ft52 + i u + i v, \ft12 + i u , 1 + i v \atop 1, 1}\right| - {\rm e}^{- 2 \tau_1}, -  {\rm e}^{- 2 \tau_2} \right)
\, , \nonumber
\end{align}
where we set $u = u_1$ and $v = v_1$. Making use of the double series representation \re{F2Series} for $F_2$, we can prove that the agreement
continues at higher order terms in the expansion as well, where each term in the double series corresponds to the $\phi (\psi_{\rm s}\bar\psi_{\rm s})^{n_1} \to \phi 
(\psi_{\rm s}\bar\psi_{\rm s})^{n_2}$  transitions. We can also confirm correctness of above predictions against explicit data of Ref.\ \cite{Bourjaily:2013mma}, 
with the first two terms being
\begin{align}
\vev{W^{[0,3,1]}_{7}}_{1,-1/2}
=
& - \frac{ {\rm e}^{- \tau_1 - \tau_2} \, {\rm e}^{\sigma_1} 
}{
(1 + {\rm e}^{2 \sigma_1}) (1 + {\rm e}^{2 \sigma_2})
( {\rm e}^{2 \sigma_1}+{\rm e}^{2 \sigma_2} + {\rm e}^{2 \sigma_1+2 \sigma_2} )}
[
{\rm e}^{2 \sigma_1} + {\rm e}^{2 \sigma_2} + 2 {\rm e}^{2 \sigma_1 + 2 \sigma_2}
]
\\
&
+
\frac{
{\rm e}^{- 3 \tau_1 - \tau_2} \, {\rm e}^{\sigma_1}}{
(1 + {\rm e}^{2 \sigma_1})^3 (1 + {\rm e}^{2 \sigma_2})^2
( {\rm e}^{2 \sigma_1}+{\rm e}^{2 \sigma_2} + {\rm e}^{2 \sigma_1+2 \sigma_2} )^3}
[
{\rm e}^{8 \sigma_1} + 2 {\rm e}^{8 \sigma_2} + 3 {\rm e}^{6 \sigma_1} + 3 {\rm e}^{6 \sigma_2} 
\nonumber\\
&
+ 
36 {\rm e}^{6 \sigma_1 + 6 \sigma_2 }
+
39 {\rm e}^{4 \sigma_1 + 4 \sigma_2} + 9 {\rm e}^{4 \sigma_1 + 2 \sigma_2} + 18 {\rm e}^{6 \sigma_1 + 2 \sigma_2} 
+ 
5 {\rm e}^{8 \sigma_1 + 2 \sigma_2} 
+ 
9 {\rm e}^{2 \sigma_1 + 4 \sigma_2} 
\nonumber\\
&
+ 42 {\rm e}^{6 \sigma_1 + 4 \sigma_2} 
+ 
10 {\rm e}^{8 \sigma_1 + 4 \sigma_2} + 24 {\rm e}^{2 \sigma_1 + 6 \sigma_2} + 51 {\rm e}^{4 \sigma_1 + 6 \sigma_2} 
+ 36 {\rm e}^{6 \sigma_1 + 6 \sigma_2}
+ 6 {\rm e}^{8 \sigma_1 + 6 \sigma_2}
\nonumber\\
&\qquad\qquad\qquad\qquad
+ 
12 {\rm e}^{2 \sigma_1 + 8 \sigma_2} + 18 {\rm e}^{4 \sigma_1 + 8 \sigma_2} + 8 {\rm e}^{6 \sigma_1 + 8 \sigma_2}
]
\nonumber\\
&\qquad\qquad\qquad\qquad
+ \dots
\, , \nonumber
\end{align}
and the rest was straightforwardly checked numerically.

\section{Two-channel NMHV conformal blocks}

Analyses along the lines spelled out in the previous section were performed for more than 2000 examples, implying that for each of the 25 different transitions in a 
given independent component of the heptagon, we constructed matrix part for up to three additional $\psi_{\rm s} \bar\psi_{\rm s}$ pairs accompanying a given parent flux-tube excitations 
within the framework of Ref.\ \cite{Belitsky:2016vyq}. This yielded the following generic expression for the NMHV conformal blocks
\begin{align}
&
\mathcal{F}^{[r_1, r_2, r_3]}_{h_1, t_1 |h_2, t_2} (u,  \tau_1 |v,  \tau_2)
\\
&\quad
=
F_2
\left.\left(
{
\frac{|h_1| + |h_2|}{2} + \frac{2 r_2 + \widehat{r}_1 + \widehat{r}_3}{4} + i u + i v \, ,
\frac{|h_1|}{2} + \frac{2 r_1 + \widehat{r}_1}{4} + i u \, ,
\frac{|h_2|}{2} + \frac{2 r_3 + \widehat{r}_3}{4} + i v
\atop
t_1 , t_2
}
\right|
- {\rm e}^{-2 \tau_1}, - {\rm e}^{-2 \tau_2}
\right)
\, , \nonumber
\end{align}
where $h_i$'s and $t_i$'s are the helicities and twists of the parent flux-tube excitations, while the hatted SU(4) labels are
\begin{align}
\widehat{r}_1 = (4 - r_1) \theta (h_1 > 0) + r_1 \theta (h_1 \leq 0)
\, , \qquad
\widehat{r}_3 = r_3 \theta (h_2 > 0) + (4 - r_3) \theta (h_2 \leq 0)
\, .
\end{align}
Then, to account for all small fermion-antifermion pairs in the independent heptagon components introduced in Sect.\ \ref{SectionIndepHeptagonComp}, one has to 
dress all integrands of the parent transitions with the above conformal blocks, i.e., 
\begin{align}
\label{Substitution}
I^{\bf\scriptscriptstyle R_1 | R_2}(\alpha_1, u |, \alpha_2, v)
\to  
I^{\bf\scriptscriptstyle R_1 | R_2}(\alpha_1, u |, \alpha_2, v) 
\mathcal{F}^{[r_1, r_2, r_3]}_{h_{\alpha_1}, t_{\alpha_1} | h_{\alpha_2}, t_{\alpha_2}}  (u,  \tau_1 |v, \tau_2)
\, .
\end{align}
Here the helicities and twists are linear functions of the $\alpha_i$ labels as can be established from explicit flux-tube content of the effective particles of 
Sect.~\ref{SectionExcitations}. Equation \re{Wr1r2r3}, with the substitution \re{Substitution}, provides the resummed NMHV heptagon.

\section{Conclusion}

In the present paper, we built fifteen independent NMHV heptagon integrands of the parent flux-tube excitations of increasing helicity. To account for all twist corrections 
accompanying these and thus to restore the exact kinematics of the tree seven-leg emplitude in conformal cross ratios, we summed over the series of small fermion-antifermion 
pairs. The latter procedure was based on the knowledge of the matrix part stemming from contraction of pentagon SU(4) tensors found by means of a recursive procedure
advocated in Ref.\ \cite{Belitsky:2016vyq} and a conjectured form of the rapidity dependence of one of its tensor structures. The outcome of this consideration was the 
construction of two-channel NMHV conformal blocks from the first several terms in their collinear expansion.

A generalization of this analysis to construction of multi-channel conformal blocks for octagon and higher polygons, either by small fermion-antifermion resummation or sl(2) 
arguments, is begging for attention. A natural space of functions is to looks for generalized supergemetric series of multiple arguments. While the exchange of a single parent 
flux-tube excitation was sufficient at tree level, at higher loops one will have to take into account two of these simultaneously. These and related questions will be addressed
in future work.

\section*{Acknowledgments}

We would like to thank Benjamin Basso and Georgios Papathanasiou for interest in the project at its initial stage. This research was supported by the U.S. 
National Science Foundation under the grant PHY-1713125.

\appendix

\section{MHV conformal block of the hexagon}
\label{ConformalBlocksSection}

The hexagon is encoded by the following reference twistors
\begin{align}
Z_1 &= (1,0,1,1)
\, , 
&
Z_2 
&
= (1,0,0,0)
\, , 
&
Z_3 
&
= (-1,0,0,1)
\, , \\
Z_4 
&= (0,1,-1,1)
\, ,
&
Z_5 
&
= (0,1,0,0)
\, , 
&
Z_6 
&
= (0,1,1,0)
\, . 
\end{align}
Notice that this construction provides a natural tessellation of null polygons: they are divided in a series of pentagon transitions that overlap on
intermediate null squares. To encode all inequivalent polygons we apply conformal symmetries of these middle squares on all twistors above
or below them. All hexagons are then defined by the set
\begin{align}
\bit{Z} = \{ Z_1 \cdot M (\tau, \sigma, \phi) \, , \  Z_2 \, , \ 
Z_3 \, ,  \ Z_4 \, , \
Z_5 \, , \  Z_6  \cdot M (\tau, \sigma, \phi) 
\}
\, .
\end{align}
Where 
\begin{align}
\label{MforRefSquare1}
M  (\tau, \sigma, \phi) = \left(
\begin{array}{cccc}
{\rm e}^{\sigma - i \varphi/2} & & & \\
& {\rm e}^{-\sigma - i \varphi/2} & & \\
& & {\rm e}^{\tau + i \varphi/2} & \\
& & & {\rm e}^{- \tau + i \varphi/2}
\end{array}
\right)
\, ,
\end{align}
is the conformal transformation leaving the intermediate square invariant
\begin{align}
Z_1 = (0,0,1,0)
\, , \qquad
Z_2 = (1,0,0,0)
\, , \qquad
Z_3 = (0,0,0,1)
\, , \qquad
Z_4 = (0,1,0,0)
\, .
\end{align}

The two twistors $Z_2$ and $Z_5$ define a channel for conformal block decomposition and there is a tower of states (primary and its
descendants) that propagate on top with dimension (actually, twist) $J$. Let us construct a representation of sl(2) generators acting on the space
of conformal cross ratios
\begin{align}
u 
&= \frac{(1234)(4561)}{(1245)(3461)} 
= {\rm e}^{2 \tau - 2 \sigma} v w
\, , \\ 
v 
&= \frac{(2345)(5612)}{(2356)(1245)} 
= \frac{1}{1 + {\rm e}^{2 \tau}}
\, , \\
w
&= \frac{(3456)(6123)}{(3461)(2356)} 
= \frac{1}{1 + {\rm e}^{2 \sigma} + 2 \cos\varphi \, {\rm e}^{\sigma - \tau} + {\rm e}^{- 2 \tau}}
\, .
\end{align}
Obviously these are invariant under an SL(4) transformation $V$, $\bit{Z} \to \bit{Z}^\prime = \bit{Z} \cdot V$, $\det V = 1$. In parallel to our discussion of 
correlation functions, we need to figure out the change of the bottom/top twistors. The side twistors determining the channel are invariant under SL(4)
transformation with unit two-by-two matrix in the left top block. So, we are left with SL(2) right bottom block,
\begin{align}
\label{SL2U}
U_\alpha (\vartheta)
=
\left(
\begin{array}{cc}
1_{[2 \times 2]} & 0 \\
0 & {\rm e}^{\vartheta \sigma_\alpha/2}
\end{array}
\right) 
\, ,
\end{align}
parametrized by $\vartheta$ and $\sigma_\alpha$ being the triplet of conventional Pauli matrices. This matrix changes the conformal frame of the bottom twistors
\begin{align}
( Z_6, Z_1) \to ( Z^\prime_6, Z^\prime_1) _\alpha = ( Z_6, Z_1) \cdot U_\alpha
\, , 
\end{align}
and defines changed cross ratios. E.g., under the $\alpha = 3$ transformation, all cross ratios change as
\begin{align}
(u,v,w) \to (u', v', w')_{\alpha = 3} = {\rm e}^{\vartheta L^0} (u,v,w)
\, , 
\end{align}
where
\begin{align}
L^0 = \ft12 \partial_\tau
\, .
\end{align}
One can find the representation of the quadratic Casimir by expanding transformed cross ratios to quadratic order in the transformation parameter $\vartheta$. One finds
\begin{align}
\sum_{\alpha = 1}^3 (u',v',w')_{\alpha} = \left( 1 + \dots + \ft{1}{2!} \vartheta^2 \mathbb{C}_{2} + \dots \right) (u,v,w)
\, ,
\end{align}
with
\begin{align}
\label{hexCasimir}
\mathbb{C}_{2} 
&= \ft12 \left( L^+ L^- + L^- L^+ \right) + (L^0)^2 
\nonumber\\
&= \ft14 {\rm e}^{-2 \tau} \left( \partial_\tau - \partial_\sigma \right)^2 + \ft14 \partial_\tau^2 + \ft12 \partial_\tau
\, .
\end{align}
The eigenvalue equation for afore-derived Casimir operator,
\begin{align}
\mathbb{C}_{2} {\rm e}^{2 i \sigma u} 
\mathcal{F}_6 (\tau, \sigma) = \ft{h}{2} (\ft{h}{2}-1){\rm e}^{2 i \sigma u} 
\mathcal{F}_6 (\tau, \sigma) 
\end{align}
where $h$ is the helicity of the intermediate excitation, gives \cite{Gaiotto:2011dt}
\begin{align}
\mathcal{F}_6 (u, \tau) 
&
=
{\rm e}^{- h \tau}
{_2 F_1} \left.\left( { \ft{h}{2} + i u ,  \ft{h}{2} + i u \atop h} \right| - {\rm e}^{-2 \tau} \right)
\\
&
+
{\rm c} \,
{\rm e}^{- (2 - h) \tau}
{_2 F_1} \left.\left( {1 - \ft{h}{2} + i u , 1 -  \ft{h}{2}+ i u\atop 2 - h} \right| - {\rm e}^{-2 \tau} \right)
\, , \nonumber
\end{align}
where we have to set ${\rm c} = 0$ to have a proper behavior as $\tau \to \infty$.

\section{MHV conformal block of the heptagon}
\label{TwistorsHeptagonSection}

The heptagon is displayed in Fig.\ \ref{Heptagon} with the corresponding reference twistors being
\begin{align}
Z_1 &= (1,0,1,1)
\, , 
&
Z_2 
&
= (1,0,0,0)
\, , 
&
Z_3 
&
= (-1,0,0,1)
\, , \qquad
Z_4 
= (-1,1,-1,3)
\, , \nonumber\\
Z_5 
&= (0,2,-1,1)
\, ,
&
Z_6
&
= (0,1,0,0)
\, , 
&
Z_7
&
= (0,1,1,0)
\, .
\end{align}
The bottom middle square is invariant under the same transformation $M$ as defined in Eq.\ \re{MforRefSquare1}, the top 
middle square is conformally invariant with respect to the matrix multiplication with
\begin{align}
\label{MforRefSquare2}
M^\prime  (\tau, \sigma, \varphi) 
= 
\left(
\begin{array}{cccc}
{\rm e}^{- \sigma - i \varphi/2} & & & - {\rm e}^{- \sigma - i \varphi/2} + {\rm e}^{\tau + i \varphi/2}  \\
& {\rm e}^{\sigma - i \varphi/2} & & \\
& {\rm e}^{\sigma - i \varphi/2} -  {\rm e}^{- \tau + i \varphi/2} &  {\rm e}^{- \tau + i \varphi/2} &  {\rm e}^{\tau + i \varphi/2} -  {\rm e}^{- \tau + i \varphi/2} \\
& & & {\rm e}^{\tau + i \varphi/2}
\end{array}
\right)
\, .
\end{align}
Then all inequivalent heptagons are parametrized by the set of twistors
\begin{align}
\label{HeptagonTwistors}
\bit{Z}
=
\{
&
Z_1 \cdot M (\tau_1, \sigma_1, \varphi_1) \, , \
Z_2 \, , \
Z_3 \, , \
Z_4 
\cdot [ M^\prime (\tau_2, \sigma_2, \varphi_2)]^{-1}
\, , \nonumber\\ 
&
Z_5 
\cdot [ M^\prime (\tau_2, \sigma_2, \varphi_2)]^{-1}
\, , \
Z_6 \, , \ 
Z_7  \cdot M (\tau_1, \sigma_1, \varphi_1) 
\}
\, .
\end{align}

The conformal cross ratios are defined in terms of these as
\begin{align}
u_1 &
= \frac{(6123)(5672)}{(5623)(6712)} = {\rm e}^{2 \tau_1}
\, , \\
v_1 &
= \frac{(5671)(6723)}{(5673)(6712)} = {\rm e}^{\sigma_1 + \tau_1 - i \varphi_1}
\, , \\
w_1&
= \frac{(1234)(6723)}{(6234)(7123)} = {\rm e}^{\sigma_1 + \tau_1 + i \varphi_1}
\, , 
\end{align}
for odd and
\begin{align}
u_2 &
= \frac{(1234)(5673)}{(5734)(7623)} = {\rm e}^{- 2 \tau_2}
\, , \\
v_2 &
= \frac{(5623)(1234)}{(5234)(6123)} = {\rm e}^{- \sigma_2 - \tau_2 - i \varphi_2}
\, , \\
w_2 &
= \frac{(4567)(5723)}{(4563)(5672)} = {\rm e}^{- \sigma_2 + \tau_2 + i \varphi_2}
\, , 
\end{align}
even invariants, respectively. 

In order to discuss symmetries of both intermediate reference squares in a uniform fashion, notice that \re{MforRefSquare2} can be brought to the diagonal
form \re{MforRefSquare1} with a transformation matrix
\begin{align}
R
=
\left(
\begin{array}{cccc}
0 & 1 & 1 & 0 \\
1 & 0 & 0 & 0  \\
1 & 0 & 1 & 1 \\
0 & 0 & 1 & 0
\end{array}
\right)
\end{align}
such that
\begin{align}
R^{-1} M^\prime (\tau, \sigma, \phi) R = M (\tau, \sigma, \phi)
\, .
\end{align}

We have now two channels, defined by the pairs of twistors $(Z_2, Z_6)$ and $(Z_3, Z_6)$. These are invariant under
SL(2) transformations
\begin{align}
U_\alpha (\vartheta_1)
\, , \qquad
R^{-1} U_\alpha (\vartheta_2) R
\, , 
\end{align}
respectively, with $U_\alpha$ introduced earlier in Eq.\ \re{SL2U}. They act on the bottom/top, respectively, twistors, i.e., 
\begin{align}
\label{botConf}
(Z_7, Z_1) 
&\to (Z^{\prime}_7, Z^{\prime}_1) = (Z_7, Z_1) \cdot U_\alpha (\vartheta_1)
\, , \\
\label{topConf}
(Z_4, Z_5) 
&\to (Z^{\prime}_4, Z^{\prime}_5) = (Z_4, Z_5) \cdot [R^{-1} U_\alpha (\vartheta_2) R]^{-1}
\, .
\end{align}

To find explicit forms of conformal Casimirs for bottom and top channels, let us substitute the above twistors into generic conformal cross ratios of the heptagon.
The latter are defined as
\begin{align}
w_i = u_{i+1, i+3, i+4, i}
\, , \qquad 
i = 1, \dots, 7
\, ,
\end{align}
with
\begin{align}
u_{i,j,k,l} = \frac{(i, i+1, j, j+1) (k, k+1, l, l+1)}{(i, i+1, k, k+1) (j, j+1, l, l+1)}
\, .
\end{align}
Then, under \re{botConf}/\re{topConf},
\begin{align}
\sum_{\alpha = 1}^3 (w'_i)_{\alpha} &= \left( 1 + \dots + \ft{1}{2!} \vartheta_1^2 \mathbb{C}_{2}^{(1)} + \dots \right) w_i
\, , \\
\sum_{\alpha = 1}^3 (w'_i)_{\alpha} &= \left( 1 + \dots + \ft{1}{2!} \vartheta_2^2 \mathbb{C}_{2}^{(2)} + \dots \right) w_i
\, .
\end{align}
we get the bottom/top conformal Casimir operators, namely,
\begin{align}
\mathbb{C}_{2}^{(1)} 
= \ft14 {\rm e}^{-2 \tau_b} \left( \partial_{\tau_b} - \partial_{\sigma_b} \right)^2 + \ft14 \partial_{\tau_b}^2 + \ft12 \partial_{\tau_b}
- \ft14 {\rm e}^{-2 \tau_b} \left( \partial_{\tau_b} - \partial_{\sigma_b} \right) \left( \partial_{\tau_t} - \partial_{\sigma_t} \right)
\, .
\end{align}
and $\mathbb{C}^{(2)}_{2}$ obtained from the above via simple substitutions. The two-channel conformal block are found as a solution to the eigenvalue 
equations for both Casimir operators simultaneously
\begin{align}
\mathbb{C}^{(1)}_{2} {\rm e}^{2 i \sigma_1 u + 2i \sigma_2 v}  \mathcal{F}_7  (u, \tau_1 | v, \tau_2) 
&
= \ft{h_1}{2} (\ft{h_1}{2} - 1) {\rm e}^{2 i \sigma_1 u + 2i \sigma_2 v}  \mathcal{F}_7   (u, \tau_1 | v, \tau_2) 
\, , \\
\mathbb{C}^{(2)}_{2} {\rm e}^{2 i \sigma_1 u + 2i \sigma_2 v}  \mathcal{F}_7   (u, \tau_1 | v, \tau_2)  
&
= \ft{h_2}{2} (\ft{h_2}{2} - 1) {\rm e}^{2 i \sigma_1 u + 2i \sigma_2 v}  \mathcal{F}_7   (u, \tau_1 | v, \tau_2) 
\, .
\end{align}
It reads \cite{Sever:2011pc}
\begin{align}
\mathcal{F}_7   (u, \tau_1 | v, \tau_2) 
&
=
{\rm e}^{- h_1 \tau_1 - h_2 \tau_2}
F_2
\left. \left( { \ft{h_1}{2} + \ft{h_2}{2} + i u + i v, \ft{h_1}{2} + i u, \ft{h_2}{2} + i v \atop h_1, h_2} \right| - {\rm e}^{- 2 \tau_1}, - {\rm e}^{- 2 \tau_2} \right)
\, ,
\end{align}
in terms of the Appell function $F_2$ determined by the infinite hypergeometric series in two variables \cite{BatErdVol1}
\begin{align}
\label{F2Series}
F_2
\left.\left(
{
a \, ,
b_1 \, ,
b_2
\atop
c_1 , c_2
}
\right|
z_1 \, , z_2
\right)
=
\sum_{n_1 = 0}^\infty
\sum_{n_2 = 0}^\infty
\frac{(a)_{n_1 + n_2} (b_1)_{n_1} (b_2)_{n_2}}{n_1! n_2! (c_1)_{n_1} (c_2)_{n_2}}
z_1^{n_1} z_2^{n_2}
\, .
\end{align}

\section{Matrix part}
\label{AppendixFermionTensor}

%%%%%%%%%%%%%%%%%%%%%%%%%%%%%%%%%%%%%%%%%%%%%%%%%%%%%%%%%%%%%%%%%%%%%
%            Figure
%%%%%%%%%%%%%%%%%%%%%%%%%%%%%%%%%%%%%%%%%%%%%%%%%%%%%%%%%%%%%%%%%%%%%
\begin{figure}[t]
\begin{center}
\mbox{
\begin{picture}(0,110)(240,0)
\put(0,-250){\insertfig{17}{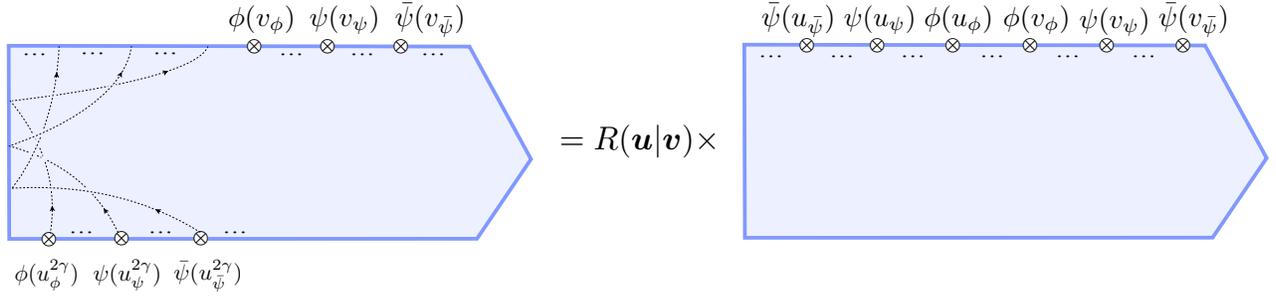}}
\end{picture}
}
\end{center}
\caption{ \label{MovingExcitationsFig} As one moves an excitation from the initial to final state, the dynamical pentagon transition acquires a
rational factor $R (\bit{u} | \bit{v})$ depending on the difference of particle rapidities between the in- and out- states.
}
\end{figure}
%%%%%%%%%%%%%%%%%%%%%%%%%%%%%%%%%%%%%%%%%%%%%%%%%%%%%%%%%%%%%%%%%%%%%

Let us spell out the rules of the game for construction of the matrix part. The minimal set of pentagon tensor structures needed to recover matrix part for all transitions 
was addressed in Ref.\ \cite{Belitsky:2016vyq}. 
\begin{itemize}
\item The rest can be easily recovered from these by moving excitations from the initial to final state. This is demonstrated 
graphically in Fig.\ \ref{MovingExcitationsFig}, where
\begin{align}
R (\bit{u} | \bit{v}) 
&
= (u_\phi | v_\phi)_1  (u_\phi | v_\phi)_2 \dots (u_\phi |v_\psi)_{3/2} \dots (u_\phi |v_{\bar\psi})_{3/2} \dots
\nonumber\\
&\times 
(u_\psi |v_\phi )_{3/2}  \dots (u_\psi |v_\psi )_1 \dots (u_\psi | v_{\bar\psi})_2 \dots
\nonumber\\
&\times 
(u_{\bar\psi} |v_\phi )_{3/2}  \dots (u_{\bar\psi} |v_\psi )_2 \dots (u_{\bar\psi}|v_{\bar\psi})_1 \dots
\, ,
\end{align}
with $(u|v)_\alpha \equiv u - v + i \alpha$.
\item The interchange of adjacent excitations in the in/out- states can be done making use of the Watson equations. 
\item All charged (or nonsinglet) pentagon transitions are recovered by taking rapidities of certain excitations to infinity. 
\end{itemize}

Let us exemplify these for a few cases  that are used in the body of the paper. We begin with transitions involving only fermions and antifermions. In the earlier paper 
\cite{Belitsky:2016vyq}, we fixed the form of all diagonal $\psi^n \to \bar\psi^n$ tensors. The simplest fermion-antifermion production form factor, i.e., $0 \to \psi \bar\psi$ 
can be then found in terms of $n=1$ transition by adopting the first rule,
\begin{align}
\label{0to2fermiontensor}
\Pi^{\bf 1}_{0 | \psi^{i} \bar\psi_{j}} (0 | u, v)
=
\frac{\delta^{i}_{j}}{(u| v)_2}
\, ,
\end{align}
where we assumed unit normalization for the transition amplitude $\Pi^{\bf 1}_{\psi^i | \bar\psi_j} (u | v) = \delta_j^i$.

The tensor for the transition $\psi \bar\psi \to \bar\psi \psi$, can be found from the one of the $\psi\psi \to \bar\psi \bar\psi$ via the relation
\begin{align}
\Pi^{\bf 1}_{\bar\psi_{j_1} \psi^{i_1} | \psi^{i_2} \bar\psi_{j_2}} (v_1, u_1 | u_2, v_2)
&=
\frac{(u_1 | u_2)_{- 1} (v_1 | v_2)_{-1} (u_1 | v_2)_0 (v_1 | u_1)_0}{(u_1 | v_1)_2 (u_1 |v_2)_0 (u_2|v_1 )_4 (u_2|v_2)_2}
\nonumber\\
&\times
[ R_{\psi\bar\psi} (u_2| v_1-2i) ]_{k_2 j_1}^{i_2 l_1}
\Pi^{\bf 1}_{\psi^{k_2} \psi^{i_1} | \bar\psi_{l_1} \bar\psi_{j_2}} (u_2 + 2i, u_1 | v_1 - 2i, v_2)
\, , 
\end{align}
where we moved all excitations from the in-state to the out-state making use of the first rule, interchanged two two middle particles by means of the Watson
equation with the fermion-antifermion R-matrix
\begin{align}
[ R_{\psi\bar\psi} (u_1| u_2) ]_{i_1 j_1}^{i_2 j_2}
=
\delta_{i_1}^{i_2} \delta_{j_1}^{j_2} + \frac{i}{(u_1|u_2)_{-2}} \delta_{i_1}^{j_2} \delta_{j_1}^{i_2}
\, ,
\end{align}
and then moving the two left-most pair back to the bottom. Using the result for $\Pi^{\bf 1}_{\psi\psi | \bar\psi\bar\psi}$,
\begin{align}
\Pi^{\bf 1}_{\psi_{i_1}\psi_{i_2} | \bar\psi^{j_1}\bar\psi^{j_2}} (u_1, u_2 | v_1, v_2)
=
\left[
1 + \frac{(u_1 | v_2)_0 (u_2 | v_1)_1}{(u_2 | u_1)_{1} (v_2 | v_1)_{1}}
\right]
\delta^{i_1}_{j_1} \delta^{i_2}_{j_2}
+
\frac{(u_1 | v_1)_0 (u_2 | v_2)_1}{(u_2 | u_1)_{1} (v_2 | v_1)_{-1}}
\delta^{i_1}_{j_2} \delta^{i_2}_{j_1}
\, ,
\end{align}
this gives us the explicit result
\begin{align}
\label{2to2fermiontensor}
\Pi^{\bf 1}_{\bar\psi_{j_1} \psi^{i_1} | \psi^{i_2} \bar\psi_{j_2}} (v_1, u_1 | u_2, v_2)
=
\delta^{i_1}_{j_2} \delta^{i_2}_{j_1} - \frac{(u_2 | v_1)_0 (u_1 | v_2 )_1}{(u_1 | v_1)_2 (u_2 | v_2 )_2} \delta^{i_1}_{j_1} \delta^{i_2}_{j_2} 
\, .
\end{align}
From this singlet transition, we can immediately derive the charged one, --- transforming in the fundamental representation of SU(4),
\begin{align}
\label{2to1fermiontensor}
[\Pi^{\bf 4}_{\bar\psi_{j_1} \psi^{i_1} | \bar\psi_{j_2}} ]^{i_2} (v_1, u_1 | v_2)
=
\lim_{u_2 \to \infty}
\Pi^{\bf 1}_{\bar\psi_{j_1} \psi^{i_1} | \Psi^{i_2} \bar\psi_{j_2}} (v_1, u_1 | u_2, v_2)
=
\delta^{i_1}_{j_2} \delta^{i_2}_{j_1} - \frac{(u_1 | v_2)_1}{(u_1 | v_1)_2} \delta^{i_1}_{j_1} \delta^{i_2}_{j_2} 
\, .
\end{align}

To produce a three-particle annihilation $\Pi^{\bf\bar{4}}_{\psi\psi\bar\psi |0}$, which is used in Sect. \ref{HeptagonExample}, we can use Eq.\ \re{2to1fermiontensor}
and move the antifermion from the final state into the initial one. The complex conjugate of this yields
\begin{align}
[\Pi^{\bf\bar{4}}_{\psi^{j_1} \psi^{j_2} \bar\psi_{j_3} |0}]_{i_3} (v_1, v_2, v_3|0)
=
-
\frac{1}{(v_1|u_2)_1 (v_2|u_2)_2} \delta^{j_2}_{j_3} \delta^{j_1}_{i_3}
+
\frac{(v_2 |u_2)_3}{(v_1|u_2)_1 (v_2|v_1)_2 (v_2|u_2)_2} \delta^{j_1}_{i_3} \delta^{j_2}_{j_3}
\end{align}

Next, we find the $[\Pi^{\bf 4}_{\bar\psi \psi | \psi \bar\psi \bar\psi}]$ tensor starting with the $\Pi^{\bf 1}_{\psi \psi \psi | \bar\psi \bar\psi \bar\psi}$. The latter is found 
via the procedure introduced in Ref. \cite{Belitsky:2016vyq} and reads
\begin{align}
\label{Pi13To3}
&
\Pi^{\bf 1}_{\psi^{i_1} \psi^{i_2} \psi^{i_3} | \bar\psi_{j_1} \bar\psi_{j_2} \bar\psi_{j_3}} (u_1, u_2, u_3 | v_1, v_2, v_3)
\\
&\qquad\qquad\qquad
= 
\delta^{i_1}_{j_1} \delta^{i_2}_{j_2} \delta^{i_3}_{j_3} \Pi_1 (u_1, u_2, u_3 | v_1, v_2, v_3)
+
\delta^{i_1}_{j_1} \delta^{i_2}_{j_3} \delta^{i_3}_{j_2} \Pi_2 (u_1, u_2, u_3 | v_1, v_2, v_3)
\nonumber\\
&\qquad\qquad\qquad
+
\delta^{i_1}_{j_2} \delta^{i_2}_{j_1} \delta^{i_3}_{j_3} \Pi_3 (u_1, u_2, u_3 | v_1, v_2, v_3)
+
\delta^{i_1}_{j_2} \delta^{i_2}_{j_3} \delta^{i_3}_{j_1} \Pi_4 (u_1, u_2, u_3 | v_1, v_2, v_3)
\nonumber\\
&\qquad\qquad\qquad
+
\delta^{i_1}_{j_3} \delta^{i_2}_{j_1} \delta^{i_3}_{j_2} \Pi_5 (u_1, u_2, u_3 | v_1, v_2, v_3)
+
\delta^{i_1}_{j_3} \delta^{i_2}_{j_2} \delta^{i_3}_{j_1} \Pi_6 (u_1, u_2, u_3 | v_1, v_2, v_3)
\, , \nonumber
\end{align}
with
\begin{align}
\Pi_6 (u_1, u_2, u_3 | v_1, v_2, v_3)
&
= 
\frac{
(u_1 | v_1)_0 (u_2 | v_1)_0 (u_1 | v_2)_0 (u_3 | v_2)_1 (u_2 | v_3)_1 (u_3 | v_3)_1
}{
( u_1 | u_2 )_{- 1} ( u_1 | u_3 )_{-1} ( u_2 | u_3 )_{- 1} ( v_1 | v_2 )_1 ( v_1 | v_3 )_1 ( v_2 | v_3 )_1
}
\, , \\
\Pi_5 (u_1, u_2, u_3 | v_1, v_2, v_3)
&
= \frac{i}{(v_1 | v_2)_0} \Pi_6 (u_1, u_2, u_3 | v_1, v_2, v_3) + \frac{(v_1 | v_2)_{-1}}{(v_1 | v_2)_0} \Pi_6 (u_1, u_2, u_3 | v_2, v_1, v_3)
\, , \nonumber\\
\Pi_4 (u_1, u_2, u_3 | v_1, v_2, v_3)
&
= \frac{i}{(v_2 | v_3)_0} \Pi_6 (u_1, u_2, u_3 | v_1, v_2, v_3) + \frac{(v_2 | v_3)_{-1}}{(v_2 | v_3)_0} \Pi_6 (u_1, u_2, u_3 | v_1, v_3, v_2)
\, , \nonumber\\
\Pi_3 (u_1, u_2, u_3 | v_1, v_2, v_3)
&
= \frac{i}{(v_2 | v_3)_0} \Pi_5 (u_1, u_2, u_3 | v_1, v_2, v_3) + \frac{(v_2 | v_3)_{-1}}{(v_2 | v_3)_0} \Pi_5 (u_1, u_2, u_3 | v_1, v_3, v_2)
\, , \nonumber\\
\Pi_2 (u_1, u_2, u_3 | v_1, v_2, v_3)
&
= \frac{i}{(v_1 | v_2)_0} \Pi_4 (u_1, u_2, u_3 | v_1, v_2, v_3) + \frac{(v_1 | v_2)_{-1}}{(v_1 | v_2)_0} \Pi_4 (u_1, u_2, u_3 | v_2, v_1, v_3)
\, , \nonumber\\
\Pi_1 (u_1, u_2, u_3 | v_1, v_2, v_3)
&
= \frac{i}{(v_1 | v_2)_0} \Pi_3 (u_1, u_2, u_3 | v_1, v_2, v_3) + \frac{(v_1 | v_2)_{-1}}{(v_1 | v_2)_0} \Pi_3 (u_1, u_2, u_3 | v_2, v_1, v_3)
\, . \nonumber
\end{align}
Then making use the fact that the small fermion at infinite rapidity is a supersymmetry transformation, we find
\begin{align}
[\Pi^{\bf 4}_{\bar\psi_{j_1} \psi^{i_2} | \psi^{i_3} \bar\psi_{j_2} \bar\psi_{j_3}}]^{i_1} (v_1, u_2 | u_3, v_2, v_3)
=
\lim_{u_1 \to \infty}
\Pi^{\bf 1}_{\bar\psi_{j_1} \psi^{i_2} \psi^{i_1} | \psi^{i_3} \bar\psi_{j_2} \bar\psi_{j_3}} (v_1, u_2, u_1 | u_3, v_2, v_3)
\, ,
\end{align}
where the tensor in the right-hand side is related via set of transformations to Eq.\ \re{Pi13To3},
\begin{align}
&
\Pi^{\bf 1}_{\bar\psi_{j_1} \psi^{i_2} \psi^{i_1} | \psi^{i_3} \bar\psi_{j_2} \bar\psi_{j_3}} (v_1, u_2, u_1 | u_3, v_2, v_3)
\\
&\qquad
=
\frac{(u_1 | u_3)_{- 1} (u_1 | v_2)_{- 1} (u_1 | v_3)_0 (u_2 | u_3)_{-1} (u_2 | v_2)_{- 1} (u_2 | v_3)_0 (v_1 | u_3)_0 (v_1 | v_2)_{- 1} (v_1 | v_3)_{- 1}
}{
(u_1 | v_1 )_2 (u_1 | v_2)_0 (u_1 | v_3)_0
(u_2 | v_1)_2 (u_2 | v_2)_0 (u_2 | v_3)_0
(u_3 | v_1)_4 (u_3 | v_2)_2 (u_3 | v_3)_2
}
\nonumber\\
&\qquad\quad
\times
[R_{\psi\bar\psi} (u_3 - v_1 + 2i)]_{k_3 j_1}^{i_3 l_1}
\Pi^{\bf 1}_{\psi^{k_3} \psi^{i_2} \psi^{i_1} | \bar\psi_{l_1} \bar\psi_{j_2} \bar\psi_{j_3}} (u_3 + 2 i, u_2, u_1 | v_1 - 2i, v_2, v_3)
\, . \nonumber
\end{align}
In this manner we deduce the explicit form
\begin{align}
[\Pi^{\bf 4}_{\bar\psi_{i_2} \psi^{i_1} | \psi^{j_3} \bar\psi_{j_2} \bar\psi_{j_1}}]^{i_3} (u_3, u_2 | v_3, v_2, v_1)
&
=
\frac{(v_1|u_2)_1}{(v_2|v_1)_{1}} \delta^{i_1}_{j_2} \delta_{i_2}^{j_3} \delta^{i_3}_{j_1}
+
\frac{(u_2|v_2)_0}{(v_2|v_1)_1} \delta^{i_1}_{j_1} \delta_{i_2}^{j_3} \delta^{i_3}_{j_2}
\\
&
+
\frac{(u_2|v_1)_1 (u_2 | v_2)_1 (v_3|u_3)_0 ( v_3 | v_1)_3}{(u_2|u_3)_2 (v_2|v_1)_1 (v_3|v_1)_{2} (v_3 | v_2 )_2} \delta^{i_1}_{i_2} \delta^{j_3}_{j_2} \delta^{i_3}_{j_1}
\nonumber\\
&
+
\frac{(v_3|u_3)_0}{(v_3|v_2)_1}
\left[
\frac{(v_2|u_2)_0}{(v_2|v_1)_1} - \frac{i (v_3|u_2)_1}{(v_3|v_1)_2 (v_3|v_2)_2}
\right]
\delta^{i_1}_{j_1} \delta_{i_2}^{i_3} \delta^{j_3}_{j_2}
\nonumber\\
&
+
\frac{(u_2|v_1)_1 (u_2|v_2)_1 (v_3|u_3)_0}{(u_2 |u_3)_2 (v_1|v_2)_{-1} (v_3|v_1)_2} \delta^{i_1}_{i_2} \delta_{j_1}^{j_3} \delta^{i_3}_{j_2}
+
\frac{(u_2|v_1)_1 (v_3|u_3)_0}{(v_2|v_1)_1 (v_3|v_1)_2} \delta^{i_1}_{j_2} \delta_{i_2}^{i_3} \delta_{j_1}^{j_3}
\, . \nonumber
\end{align}

As a final demonstration, let us now add scalars to the mix. As a case of study, we will find the $\Pi^{\scriptstyle\bf 6}_{0|\psi\bar\psi \phi}$ charged creation form factor 
transforming in the {\bf 6} of SU(4). The starting point of this consideration is the singlet $\Pi^{\bf 1}_{\psi^{i_1} \phi^{i_2 i_3} | \bar\psi_{j_1} \bar\psi_{j_2} \bar\psi_{j_3}}$ 
transition \cite{Belitsky:2016vyq}
\begin{align}
\Pi^{\bf 1}_{\psi^{i_1} \phi^{i_2 i_3} | \bar\psi_{j_1} \bar\psi_{j_2} \bar\psi_{j_3}} (u_1, u_2 |v_1, v_2, v_3)
&
= \delta^{i_1}_{j_1} \left( \delta^{i_2}_{j_2} \delta^{i_3}_{j_3} -  \delta^{i_3}_{j_2} \delta^{i_2}_{j_3} \right) \Pi_1  (u_1, u_2 |v_1, v_2, v_3)
\nonumber\\
&
+ \delta^{i_1}_{j_2} \left( \delta^{i_2}_{j_1} \delta^{i_3}_{j_3} -  \delta^{i_3}_{j_1} \delta^{i_2}_{j_3} \right) \Pi_2  (u_1, u_2 |v_1, v_2, v_3)
\nonumber\\
&
+ \delta^{i_1}_{j_3} \left( \delta^{i_2}_{j_1} \delta^{i_3}_{j_2} -  \delta^{i_3}_{j_2} \delta^{i_2}_{j_1} \right) \Pi_3  (u_1, u_2 |v_1, v_2, v_3)
\, ,
\end{align}
with
\begin{align}
\Pi_3 (u_1, u_2 | v_1, v_2, v_3)
&
= 
\frac{
(u_1 | v_1)_0 (u_1 | v_2)_0 (u_2 | v_3)_{3/2}
}{
( u_1 | u_2 )_{- 3/2} ( v_1 | v_2 )_1 ( v_1 | v_3 )_1 ( v_2 | v_3 )_1
}
\, , \\
\Pi_2 (u_1, u_2 | v_1, v_2, v_3)
&
= \frac{i}{(v_2 | v_3)_0} \Pi_3 (u_1, u_2 | v_1, v_2, v_3) + \frac{(v_2 | v_3)_{-1}}{(v_2 | v_3)_0} \Pi_3 (u_1, u_2 | v_1, v_3, v_2)
\, , \nonumber\\
\Pi_1 (u_1, u_2 | v_1, v_2, v_3)
&
= \frac{i}{(v_1 | v_2)_0} \Pi_2 (u_1, u_2 | v_1, v_2, v_3) + \frac{(v_1 | v_2)_{-1}}{(v_1 | v_2)_0} \Pi_2 (u_1, u_2 | v_2, v_1, v_3)
\, , \nonumber
\end{align}

The sextet pentagon can be found from this by sending the rapidities of the two antifermions to infinity, i.e., 
\begin{align}
[ \Pi^{\scriptstyle\bf 6}_{\psi^{i_1} \phi^{i_2 i_3} | \bar\psi_{j_1}}]_{j_2 j_3} (u_1, u_2 |v_1)
=
\lim\limits_{v \to \infty}
v^2 \Pi^{\bf 1}_{\psi^{i_1} \phi^{i_2 i_3} | \bar\psi_{j_1} \bar\psi_{j_2} \bar\psi_{j_3}} (u_1, u_2 |v_1, z v, \bar{z} v)
\, ,
\end{align}
where $z + \bar{z} = 1$. Next, moving the scalar and the fermion to the final state, we find the tensor for the production form factor
\begin{align}
[\Pi^{\scriptstyle\bf 6}_{0|  \phi^{i_2 i_3} \psi^{i_1}\bar\psi_{j_1}}]_{j_2 j_3} (0| u_2, u_1, v_1)
=
\frac{1}{(u_2|v_1)_{3/2} (u_1|v_1)_{2}}
[ \Pi^{\scriptstyle\bf 6}_{\psi^{i_1} \phi^{i_2 i_3} | \bar\psi_{j_1}}]_{j_2 j_3} (u_1 + 2i, u_2 + 2i |v_1)
\, ,
\end{align}
explicitly,
\begin{align}
[\Pi^{\scriptstyle\bf 6}_{0| \phi^{i_2 i_3} \psi^{i_1}\bar\psi_{j_1}}]_{j_2 j_3} (0| u_2, u_1, v_1)
&=
\frac{(u_2|v_1)_{7/2}}{(u_1|u_2)_{-3/2} (u_2|v_1)_{3/2} (u_1|v_1)_2}
\delta^{i_1}_{j_1} ( \delta^{i_2}_{j_2} \delta^{i_3}_{j_3} - \delta^{i_2}_{j_3} \delta^{i_3}_{j_2})
\nonumber\\
+
\frac{1}{(u_2|u_1)_{3/2} (u_2|v_1)_{3/2}}
&
\left[
(\delta^{i_1}_{j_3} ( \delta^{i_2}_{j_2} \delta^{i_3}_{j_1}  - \delta^{i_2}_{j_1} \delta^{i_3}_{j_2}) 
+ 
\delta^{i_1}_{j_2} ( \delta^{i_2}_{j_1} \delta^{i_3}_{j_3} - \delta^{i_2}_{j_3} \delta^{i_3}_{j_1} ) 
\right]
\, .
\end{align}

%%%%%%%%%%%%%%%%%%%%%%%%%%%%%%%%%%%%%%%%%%%%%%%%%%%%%%%%%%%%%%%%

%%%%%%  Bibliography %%%%%%%%%%%%%%%%%%%%%%%%%%%%%%%%%%%%%%%%%%%%

\end{document}